\documentclass[journal]{IEEEtai}

\usepackage[colorlinks,urlcolor=blue,linkcolor=blue,citecolor=blue]{hyperref}
\usepackage{color,array}
\usepackage{graphicx}
\usepackage{enumerate}

\usepackage{lscape}
\usepackage{lipsum}
\usepackage{tabularx}
\usepackage{multirow}
\usepackage{longtable}
\usepackage{braket}
\usepackage{tikz}
\usepackage{tcolorbox}
\usepackage{amsmath}
\usepackage{amssymb}
\def\checkmark{\tikz\fill[scale=0.4](0,.35) -- (.25,0) -- (1,.7) -- (.25,.15) -- cycle;}

\begin{document}
%%\title{Quantum Computing Toolkit: A Sack of Quantum Tools with nuts and bolts} 
\title{Quantum Computing Toolkit\\From Nuts and Bolts to Sack of Tools} 

\author{Himanshu Sahu\textsuperscript{1}, \IEEEmembership{Student Member, IEEE}, Hari Prabhat Gupta\textsuperscript{2} \IEEEmembership{Senior Member, IEEE}
\thanks{\textsuperscript{1}Himanshu Sahu is working as a research scholar in Computer Science and Engineering Department, Indian Institute of Technology (BHU), Varanasi. email: \url{himanshusahu.rs.cse21@iitbhu.ac.in}.\\
\textsuperscript{2} HariPrabhat Gupta is with Computer Science and Engineering Department, Indian Institute of Technology (BHU), Varanasi as an Associate Professor. email: \url{hariprabhat.cse@iitbhu.ac.in}.}\\
}

\markboth{, Vol. vv, No. n, Jan 2023}
{Sahu  H. \MakeLowercase{\textit{et al.}}: Quantum Computing Toolkit:  From Nuts and Bolts to Sack of Tools}
\maketitle
\thispagestyle{empty}

\begin{abstract}
%%hsahu
%% The word count is ~250%%

Quantum computing has the potential to provide exponential performance benefits in processing over classical computing. It utilizes quantum mechanics phenomena (such as superposition, entanglement, and interference) to solve a computational problem. It can explore atypical patterns over data that classical computers can't perform efficiently. Quantum computers are in the nascent stage of development and are noisy due to decoherence, \textit{i.e.}, quantum bits deteriorate with environmental interactions. It will take a long time for quantum computers to achieve fault tolerance although quantum algorithms can be developed in advance. Heavy investment in developing quantum hardware, software development kits, and simulators has led to a multiplicity of quantum development tools. 
Selection of a suitable development platform requires a proper understanding of the capabilities and limitations of these tools. Although a comprehensive comparison of the different quantum development tools would be of great value, to the best of our knowledge, no such extensive study is currently available.
% Comparing various quantum development tools in a comprehensive manner will be valuable, however, to the best of our knowledge, there is currently no such comprehensive study available. 
% \textit{A contrasting study of quantum development tools will be really beneficial but in best of our knowledge there in no such comprehensive article with this approach.}  
So, our paper will provide a thorough description of these tools and compare them in terms of their utility, capacity, cost, efficiency, and community support. It will also provide guidelines for using the given tools and an end-to-end tutorial for designing a quantum solution for a combinatorial optimization problem.  
\end{abstract}

% \begin{IEEEImpStatement}
% Quantum Computing has the potential not only to improve the classical algorithm's performance but also solve a set of problems that are intractable by classical computers. A theoretical basis supports quantum supremacy, although practical solutions are required along with the implementation tools for its realization. The present survey has been written in a lucid manner to provide the details of all available quantum development platforms. The paper objectively compares different platforms based on usefulness, accessibility, capacity, and cost. A tutorial will also help to guide a sample algorithm implementation on various tools. So, this paper will help in learning quantum computing for newcomers and make them capable of developing and debugging quantum algorithms. 
% \end{IEEEImpStatement}


\section{Introduction}
%%When Physics meets computing
\IEEEPARstart{Q}{uantum} Computing (QC) is a disruptive computing technology based on bewildering though prepossessing principles of Quantum Mechanics (QM) \cite{dirac1981principles}. QC envision is jointly credited to Paul Benioff and Richard Feynman, as Benioff provided a turing model of computer based on QM \cite{benioff1980computer} and Feynman in his talk motivated to simulate QM phenomenon using computers \cite{feynman2018simulating}. Afterwards extensive research has proved that QC is capable of providing efficient and pragmatic solutions to different domains, such as security \cite{gisin2002quantum}, finance \cite{orus2019quantum}, medicine \cite{cao2018potential}, communication  \cite{gisin2007quantum}, and sciences \cite{cao2019quantum}. 
QC is next frontier of computing, offering the possibility of surpassing current computing models, including Von Neumann, post-Von Neumann, and non-Von Neumann computing, with the advent of fault-tolerant quantum computers \cite{shor1996fault} and highly efficient quantum algorithms \cite{Hidary2021}.

% QC is a future computing continuum which consists the potential to surpass all existing computing paradigms (including Von Neumann, post-Von Neumann, and non-Von Neumann computing) with the support of fault-tolerant computers \cite{shor1996fault} and efficient quantum algorithms \cite{Hidary2021}.

 The QM revolutionized the world by redefining Newtonian physics and perspective toward the universe \cite{dirac1981principles}. Its seemingly mystical principles, such as entanglement, interference, and superposition, has demystified previously unknown phenomenon such as dual wave-particle behaviour \cite{messiah2014quantum}. It has become the foundation of not only quantum physics but many different fields such as quantum chemistry \cite{levine2009quantum}, quantum information theory \cite{Nielsen2002}, quantum cryptography \cite{bennett1992quantum}, Quantum Machine Learning (QML) \cite{biamonte2017quantum} and quantum sensing \cite{Djordjevic2022}.

Quantum supremacy (QS) \cite{boixo2018characterizing} refers to exponential performance speedup provided by the QC over Classical Computing (CC). The QS is proved on a theoretical and mathematical basis and also verified experimentally by Google \cite{arute2019quantum} using its sycamore processor.  
% \footnote{ After three years, researchers  have challenged this by creating equivalent performance as sycamore processor but it doesn't denies the QS} . It states that QC can efficiently solve a set of valuable problems that CC fails to solve. 
QC has taken real shape with the help of material science, quantum physics, and marvellous engineering. Still, these devices are noisy with a limited number of quantum bits (or qubits) and no error correction. Present-day QC devices are termed Noisy Intermediate Scale Quantum (NISQ) era devices since different qubit implementations (such as photonic or spin qubit), suffer temporal instability of quantum states. The environmental interactions via the qubit read-write interface introduces noise to the quantum system which causes qubit decoherence. Therefore, the quantum solutions need to be very well-designed with optimized usage of qubits, fewer errors, and noise handling technique to accomplish QS.
With QS expectations, many industry giants (Google, Microsoft, IBM) and new players (Xanadu, Rigetti) came into the picture for the quantum development race. IBM and Google in their respective quantum development road maps \cite{QJIBM} and \cite{QJGoogle}, show significant developments in the QC technologies (hardware and software) and strong future prospects. Initiated with IBM cloud \cite{Qiskit}, the QC are now available in the form of Quantum Cloud Computers (QCC) via service providers as \cite{braket, azure}. Still, the availability of QCC is limited in terms of access cost, device capacity (10-100 qubits) and run-time (long queues). Alternatively, quantum development can be done via simulation over CC. Quantum simulation is useful in developing accurate algorithms due to its noise-free behaviour however simulations are practically limited with qubit capacity due to exponentially growing memory requirements. Currently, a number of quantum hardware (Google-Sycamore, IBM-Q53, Eagle, Xanadu-X24, Rigetti-19Q Acorn), as well as software development kit (IBM-Qiskit, Google-Cirq, Xanadu-Pennylane, Microsoft-Q\# ), are available which provides simulation, emulation, hardware access and cross-platform development.  
% Upto 32 qubits simulations are feasible with an Intel CPU (8-core) and 8 GB of RAM. 

The QC advantages attract researchers from different domains (such as quantum physics, computer science, and electrical engineering) to pursue the new field and significantly contribute to quantum development. Consequently, many new entrants start their quantum journey with different backgrounds and expertise levels. So a newcomer can easily be trapped in the multiplicity of quantum development paths and guiding principles will be required to move ahead swiftly. 
% Currently, many development paths are available, which is advantageous but can also be misleading for newcomers. 
% The paper has been handcrafted to provide a solution so that a  naive user or a developer can easily make a suitable choice. 
Therefore, our paper is compiled in such a way that a naive quantum developer can create a strong understanding of QC concepts and step ahead for quantum development. With the help of our paper, a suitable choice of development platform can be easily made. 
% To achieve this we reviewed majority of quantum development tools and their critical comparison with an end-to-end tutorial on quantum development. 

\begin{table}[ht]
\caption{List of Abbreviations}
 \label{tab:1}%
\centering
\begin{tabular}{m{2.2cm}m{5.5cm}}
\hline
\textbf{Abbreviation} & \textbf{Definition}\\
\hline
AA & Amplitude Amplification\\
AQ & Algorithmic Qubit\\
AWS & Amazon Web Services\\
CC & Classical Computing/Classical Computer\\
CLOPS & Circuit Layer Oper ions per second\\
CPTP & Completely Positive Trace-preserving \\
DM & Density Matrix\\
EPR  & Einstein–Podolsky–Rosen\\ %% NOT USED YET
GCP & Google Cloud Platform\\
HHL & Harrow-Hassidim-Lloyd \\
LA & Linear Algebra\\
NISQ & Noisy Intermediate Scale Quantum \\
PQC & Parameterized Quantum Circuit\\
OQC & Oxford Quantum circuits\\
QCC & Quantum Cloud Coter\\
QC & Quantum Computing/Quantum Computers \\
QDK & Quantum Development Kit\\
QDL & Quantum Deep Learning\\
QDLC & Quantum Development Life Cycle \\
QEC & Quantum Error Correction\\
QFT & Quantum Fourier Transform\\
Qiskit & Quantum Information Science kit\\
QML & Quantum Machine Learning\\
QM & Quantum Mechanics \\
QAOA & Quantum Approximate Optimization Algorithm \\
qRAM & Quantum Random Access Memory\\
QPS & Quantum programming studio\\
QS & Quantum Supremacy\\
Qubit & Quantum Bits\\
QuEST & Quantum Exact Simulation Toolkit \\
QV & Quantum Volume\\
QVM & Quantum Virtual Machine\\
SV & State Vector\\
SC & Super Conductive\\
SSQ  & Squeezed State Qubits\\
TFQ & Tensor FLow Quantum\\
TI & Trapped Ion\\
TN & Tensor Network\\
VQE/VQA & Variational Quantum Eigensolver/Algorithm\\
\hline
\end{tabular}
\end{table}

% It will also cover the underlying fundamentals and other prerequisites.
Even though survey papers exist in the field of QC as illustrated in Table \ref{tab:survey} but the scope of the paper is either limited to surveying quantum hardware technology \cite{gyongyosi2019survey} or QC development \cite{adedoyin2018quantum}. 
% Even though there are various existing surveys available in the field of QC, their scope is either limited to surveying quantum hardware technologies\cite{gyongyosi2019survey} or QC theories and development\cite{adedoyin2018quantum}.
Some recent surveys cover the field of QML and its applications \cite{massoli2022leap}\cite{ramezani2020machine}. The authors in \cite{upama2022evolution, serrano2022quantum, garhwal2021quantum} surveyed a similar field as proposed in our paper but none of them is comprehensive and detailed enough to create a complete landscape. In order to fulfil this gap, our paper will cover prominent quantum development tools and their comparative analysis, Quantum Development Life Cycle (QDLC) and an illustrative tutorial. 
% Some authors also surveyed in the similar direction as our paper i.e., Quantum development platforms as in \cite{upama2022evolution}\cite{serrano2022quantum}\cite{garhwal2021quantum}. As compared to the previous surveys, present paper is different in the sense that it is a survey cum tutorial. It also thoroughly compares existing development platforms, including a pathway to develop an application across platforms to compare their viability. 
% A detailed comparison of the existing surveys with their limitations highlighted in contrast to our review is provided in Table \ref{tab:survey}. 
The contributions of our paper are given as follows:
\begin{enumerate}
    \item We summarized the required nuts and bolts such as underlying QM principles, QC components and 5-layer QC architecture to create the foundation of quantum development.
    \item Next, we conducted a thorough survey of quantum algorithms, provided a classification along with complexity classes and discussion on standard quantum algorithms. 
    \item The key contribution is the detailed review of prominent quantum development tools including the simulator and QC. We believe that none of the existing surveys has provided a detailed discussion and comparative analysis. 
    \item Next, we abstracted a QDLC which is more relevant to quantum software development as compared to traditional software development life cycle and their quantum derivatives.
    % provided in \cite{dey2020qdlc,weder2022quantum}. 
    \item We also described how these building blocks are developed in three tools (Cirq, Qiskit and Pennylane) and a ``Hello quantum world" example. At the end, we also provided an end-to-end tutorial using Qiskit and Pennylane which helps in learning to write a quantum code from scratch.  
\end{enumerate}

% \begin{enumerate}[i).]
%     \item A brief introduction to quantum fundamentals, including linear algebra. 
%     \item A comparative study of Quantum Development Platform with their features and limitations.
%     \item A Design choice selection guide for newcomers in the field of quantum computing.
%     \item An end-to-end quantum tutorial to understand the implementation aspects of the different development platforms. 
%     % \item State of Art solutions comparisons on the development platform and their utility.
% \end{enumerate}

\begin{table}[hb]
\caption{Summary of Recent Surveys in the field of Quantum Computing  }
 \label{tab:survey}%
\centering
\begin{tabular}{m{1.5cm}m{2cm}m{4cm}}
\hline
Publication & Scope & Limitations   \\
\hline
Laszlo Gyongyosi et. al \cite{gyongyosi2019survey}& quantum hardware  & Only hardware and its related technologies have been discussed with algorithm implementation.\\
Abhijith J. et al. \cite{adedoyin2018quantum}   & quantum algorithms & Implemented 20 algorithms but considered only IBM platform. \\
Fabio Valerio Massoli et al. \cite{massoli2022leap} &  quantum machine learning & Provided a detailed discussion of QML but no discussion of implementation or tools.\\
Paramita Basak Upama et. al. \cite{upama2022evolution}  & quantum development tools & Survey is not exhaustive, and only brief discussion has been provided, no end-to-end tutorial.\\
Maneul A. Serrano et al.\cite{serrano2022quantum}& quantum development tools &  Provided a comparison of different development tools and their qualitative assessment, but lacks how the tools can be utilized for quantum development. \\
S. Garhwal et al. \cite{garhwal2021quantum}  & quantum programming languages & Provided thorough discussion of Quantum programming languages but no discussion on the Development platform or Quantum Cloud computer. \\
Somayeh Bakhtiari Ramezani et al.\cite{ramezani2020machine}& quantum machine learning &  Discussed different QML algorithms but no discussion of implementation. \\

Özlem Salehi et al. \cite{salehi2022computer}  & quantum learning & Provided learning methodology of QC with curriculum perspective but no discussion of tools and techniques. \\
\hline
\end{tabular}
\end{table}
To deliver the contributions mentioned, the paper is organized as follows. The fundamentals are discussed in Section \ref{sec:2} and the paradigm of quantum development and quantum algorithms are provided in Section \ref{sec:3}. Section \ref{sec:4} provides details of quantum development platforms, including QCC and simulation tools and Section \ref{sec:5} explains the QDLC with an end-to-end tutorial, and finally,  Section \ref{sec:6} provides the conclusion.

% To deliver the contributions as mentioned above, the paper is organized as follows. The fundamentals are discussed in Section \ref{sec:2}, including QM principles, QC building blocks, and linear algebra. Section \ref{sec:3} discusses the paradigm of quantum development and quantum algorithms, and section \ref{sec:4} provides details of quantum development platforms, including quantum cloud computers and simulation tools. Section \ref{sec:6} explains the quantum development lifecycle with an end-to-end tutorial, and finally,  Section \ref{sec:7} provides the discussion and the conclusion.
\section{Nuts and Bolts: Fundamentals and Preliminaries of Quantum Computing}
\label{sec:2}
% The first step in the quantum journey is to understand the building blocks of the QC, referred to here as the Nuts and bolts.
This section lays the foundation for quantum development by providing key QM principles and corresponding computing counterparts, QC fundamentals and components, such as qubits, quantum logic gates, and circuits.
Also, a brief discussion on linear algebra is provided that is necessary to understand the functionalities of QC.  
% The mathematical understanding used to describe and manipulate all these principles is a branch of mathematics,\textit{ i.e.,} linear algebra (LA). So, at the end of this section, LA is discussed, covering all the essentials for QC. 
\subsection{Quantum Mechanics: The Underlying Phenomenon}
The QM is a successor of classical mechanics, that provides a more precise representation of a quantum system, \textit{i.e.}, a system described at atomic and subatomic levels. Classical mechanics fails to describe phenomena (such as double slit experiments and the dual nature of photons) that can be explained using QM theories. In a quantum system, a particle can be in a superposition state (\textit{i.e}., more than one state at the same time) and can create entanglement with other particles that can be maintained without bothering about the distance. The limitation is that a quantum state can't be cloned, and when observed, it losses its superposition (or entanglement) state and falls to a definite state. The entire quantum system is described with a set of QM principles as superposition, entanglement, no-cloning theorem and interference.  Table \ref{tab:QMP} provides the relevance of these QM principles and their applications in QC. Basic phenomena and principles which lays the foundation of QC are discussed as follows: 
%% Incomplete 
%%%Define a quantum mechanical system before defining its phenomenon.  
%With the quantum revolution in physics and motivation from Richard Feynman's lecture, the same principles are investigated to change the world of computation. These QM principles are discussed below: 
    \begin{itemize}
        \item Superposition: It is one of the key principles of QM which says that a quantum state can be represented as a sum of two or more distinct quantum states. In other words, two or more quantum states can be added together,\textit{ i.e., } “superimposed,” to create a valid quantum state. In a superposition state, a particle can be in more than one state at the same time. Observing the state of a particle will probabilistically end up in a definite state. 
        \item Entanglement: It is a quantum physical phenomenon that entangles or groups two or more particles in such a way that the state of a particle can't be described independently. The sustainability of entanglement is independent of the distance between the particles. The behavior of an entangled pair (or group) will be the same on applying QM functions, and on observing, both (or all) end up in the same definite state.  
        \item Interference: The wave functions of two particles can superimpose in-phase or out-of-phase, where in-phase interaction is known as superposition and out-of-phase interaction is known as interference. Using interference, a quantum state can be controlled by reinforcing or diminishing the waveform of a particle. 
        \item Measurement postulates: It states that a particle remains in the superposition state until observed; after measurement, it will produce one of the two possible states with the probability as per the probability amplitudes.
        \item No-cloning theorem: It states that a quantum state can not be stored or recreated that is independent and identical to an arbitrary quantum state. It is derived from the measurement postulate which says that a quantum state can't be measured without destroying the superposition. 
    \end{itemize}
% These QM principles seem representative of a quantum system in which a photon can remain in a superposition state (set of states at the same time). It can create entanglement with a different photon which can be maintained without bothering about the distance. The limitation is that it can't be cloned, and when observed, it losses its superposition state and falls to a definite state with a probability. Now, what is the relevance of these QM principles, and how can these be explored in computing and other paradigms? The answer lies in Table \ref{tab:QMP}, which describes the application of each QM principle in QC. 
%% Add a table 
\begin{table}[ht]
\caption{Applications of Quantum Mechanical principles in Quantum computing}
 \label{tab:QMP}%
\centering
\begin{tabular}{m{1.5cm}m{6cm}}
\hline
QM Principle & Quantum Computing Application   \\
\hline
Superposition & It creates a basis of a quantum bit that can reside in the superposition of $\ket{0}$ and $\ket{1}$ at the same time.\\ \hline
Entanglement & It is used in quantum teleportation and super dense coding.  \\ \hline
Interference & Interference is used to affect the probability amplitudes and thereby reach to solution in QC. (e.g. Grover search) \\ \hline
Measurement & To yield a real-world result, a qubit is measured, which produces a classical bit. \\ \hline
No cloning theorem & Hinders the traditional error correction methods on qubits. It also blocks the creation of traditional repeaters. \\\hline
\end{tabular}
\end{table}

These QM principles are used to build QC components, such as qubits and logic gates, which will be discussed in the next subsection. 

\subsection{Quantum Nuts: The Building Blocks of QC} 
The QC theory revolves around QM principles and operates on qubits. A bit is a basic unit of information in CC, which can store a value of either 0 or 1. A qubit, on the other hand, is the basic unit of information in QC, which can exist in a superposition of basis states ($\ket{0}$ and $\ket{1}$). A qubit can be manipulated using quantum gates which change a qubit state. Quantum algorithms are implemented using quantum circuits, where a quantum circuit is a series of quantum gates that manipulates qubits and provide results in classical bits. The subsection delves into the details of these building blocks.

% QC are realized as quantum circuits, in which the quantum gates are deployed to implement the unitary operations on qubits, and classical measurements are used to obtain the output. 
% Apart from the Universal gate model, Quantum annealers are present, which work on continuous variable QC and qumodes instead of qubits.
 
\paragraph{Qubits}
A qubit is based on the superposition principle by which it can store intermediate values between 0 and 1. The state of a qubit $\ket{\psi}$, can be one of the basis state ($\ket{0}$ or $\ket{1}$) or the superposition (linear combination) of ($\ket{0}$ and $\ket{1}$) as shown in Eq. \eqref{eq:qubit}.
\begin{equation}
\label{eq:qubit}
\ket{\psi} = \alpha \ket{0}  + \beta\ket{1}, 
\end{equation}
where, $\ket{\psi}$ is a unit vector in two-dimensional complex Hilbert space $\mathbb{H}$, and $\alpha,\beta$ are complex numbers that represent the probability amplitudes for measurement and $|\alpha|^2 + |\beta|^2 = 1 $. When we measure a qubit, the output will be $\ket{0}$ with probability of $|\alpha|$, and $\ket{1}$ with probability of $|\beta|$. The basis vectors $\ket{0}$ and $\ket{1}$ represented as Eq. \eqref{eq:column}.
 \begin{align}
 \label{eq:column}
 \ket{0} &= \begin{bmatrix}
           1 \\
           0 \\ 
           \end{bmatrix}, \hspace{4cm}
\ket{1} &= \begin{bmatrix}
           0 \\
           1 \\
           \end{bmatrix}             
 \end{align}
\\
Other than the standard basis($\ket{0}, \ket{1}$), the Hadamard basis can also be used to perform measurements, state tomography, and other quantum information processing tasks. It is a set of mutually unbiased bases, plus ($\ket{+})$ and minus ($\ket{-}$) given by Eq. \eqref{eq:plus}. 
 
 \begin{align}
 \label{eq:plus}
 \ket{+} &= \begin{bmatrix}
           \frac{1}{\sqrt(2)}\\
           \frac{1}{\sqrt(2)}\\
           \end{bmatrix}, \hspace{2.8cm}
\ket{-} &= \begin{bmatrix}
           \frac{1}{\sqrt(2)}\\
           -\frac{1}{\sqrt(2)} \\
           \end{bmatrix}             
 \end{align}
\\
 
\begin{figure}[ht]
  \centering
  \includegraphics[scale=0.3]{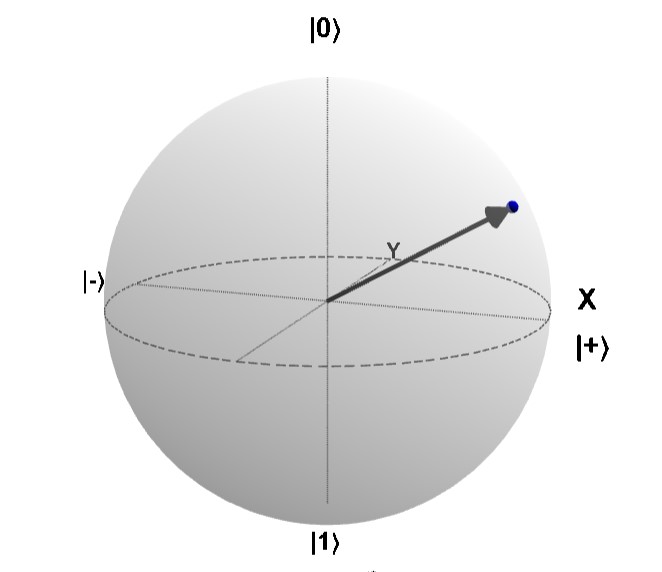}
  \caption{A Bloch sphere representation of qubit $\ket{\psi}$ = $\sqrt{0.75}\ket{0}$+$\sqrt{0.25}e^{i^{1.95\pi}}\ket{1}$. The image is drawn using the tool available at \url{https://javafxpert.github.io/grok-bloch/}}
  \label{fig:bloch}
\end{figure}

 The state of a qubit $\ket{\psi}$ is represented by braket notation utilizing the bra($\bra{}$) and ket($\ket{}$) operators where, $\bra{\psi}$ notation is for row vector and $\ket{\psi}$ notation is used for column vector. A Bloch sphere is a visual representation of a qubit in a unit sphere as shown in Fig. \ref{fig:bloch}. A qubit representation as in Eq. \eqref{eq:qubitrad} forms the basis of the Bloch sphere representation of the qubit. 
 
\begin{equation}
\label{eq:qubitrad}
|\psi\rangle = \cos{\tfrac{\theta}{2}}|0\rangle + e^{i\phi}\sin{\tfrac{\theta}{2}}|1\rangle    
\end{equation}
where, $\theta$ and $\phi$ are angle made by $\ket{\psi}$ with Z-axis and X-axis respectively.
Several interactive tools are available for Bloch sphere visualization as given in \cite{liao2022interactive}.

\begin{figure}
    \centering
    \includegraphics[width=\linewidth]{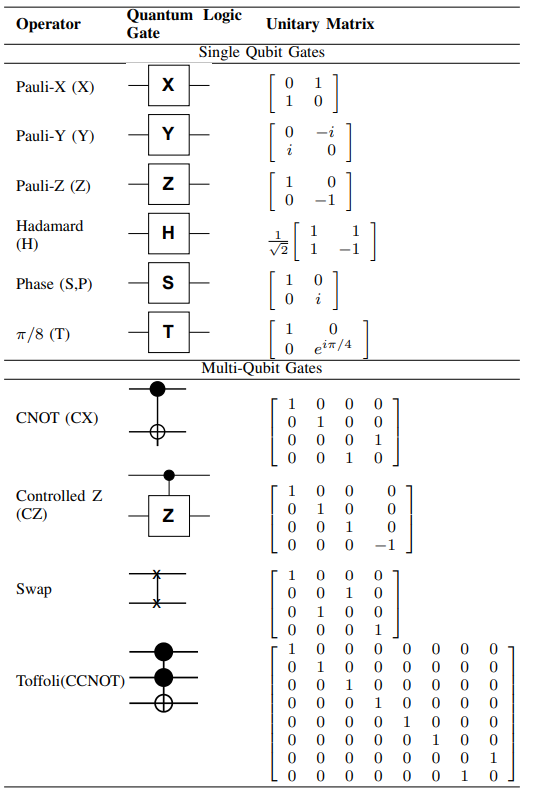}
    \caption{Quantum Logic Gates}
    \label{fig:Qgates}
\end{figure}

% Fig. \ref{fig:bloch} shows a Bloch sphere for the quantum state $\psi$ = $\sqrt{0.75}\ket{0}$+$\sqrt{0.25}e^{i^{1.95\pi}}\ket{1}$. \\
\paragraph{Quantum Gate} The quantum logic gates or quantum gates \cite{barenco1995elementary} are used to apply the quantum operations on qubits. The quantum gates are the realization of the unitary matrix where a unitary matrix $U$, is defined as a square matrix with $U^{\dagger}$ = $U^{-1}$, the inverse of a unitary matrix is the same as its conjugate transpose($\dagger$). In contrast with CC, quantum gates are reversible where the input can be obtained using the output. The list of standard quantum gates is shown in Fig. \ref{fig:Qgates}. A logic gate can be applied on single or multiple qubits. Pauli-(X, Y, Z) and Hadamard gates are examples of single-qubit gates. CNOT, SWAP, and Toffoli gates are examples of multi-qubit gates.
A quantum gate is simply realized by the matrix multiplication to the qubit state. For example, the operation of Pauli-X gate on qubit (defined in Eq.\eqref{eq:qubit}) is shown as follows:\\
\begin{equation}
    \ket{\psi}^{'} =  {\left[\begin{array}{cc} 0&1 \\1&0 \end{array}\right]}{\left[\begin{array}{c}\alpha \\ \beta \end{array}\right]} = \beta\ket{0}+\alpha\ket{1}.
\end{equation}

Universal set of quantum gates \cite{divincenzo1995two} are also defined which provides a representation of any quantum operation in the form of universal gates. The Clifford gate (CNOT, Hadamard, Swap) and Toffoli gates is one of the universal gates set. Another set of universal gates contains a rotation gate, phase shift, and CNOT gate. 

\begin{figure}[t]
    \centering
    \includegraphics[scale=0.4]{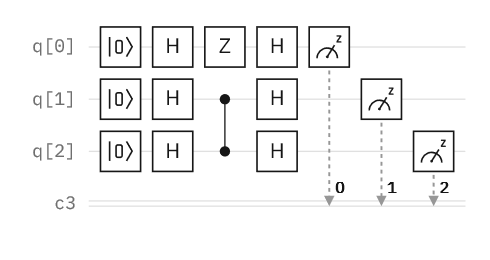}
    \caption{A Representative Quantum Circuit with 3 input qubits, containing Hadamard gate and Flip gate and 3 classical bits for measurement The following circuit is created using the \url{https://quantum-computing.ibm.com/composer/}} 
     \label{fig:Qcir}
    \end{figure}

\paragraph{Quantum Circuit} A Quantum circuit is a linear arrangement of quantum logic gates which represents a quantum solution or implementation of a quantum algorithm. The input to the quantum circuit is one or more qubits on which unitary transformations are applied via quantum gates. The sequence of logic gates transforms the state of the qubit and at the end, a measurement gate is applied to output in the form of a classical bit. A quantum circuit is shown in Fig. \ref{fig:Qcir} which shows three qubits as inputs and application of Hadamard, swap and Pauli-z gates followed by measurement.

\paragraph{Quantum Memory} A quantum memory or quantum random access memory (qRAM) \cite{giovannetti2008quantum} is a device which is capable of storing the qubit state without decoherence.
When a quantum particle interacts with the environment it changes its state which is termed as decoherence. Similarly, a qubit realized on a QC may change its state due to memory access interface. So, a qRAM is designed in such a way that it can prevent the decoherence of the qubit for some significant amount of time. Quantum memory doesn't violate the no-cloning theorem if all qubits are distinguishable. The function of a qRAM is mathematically described as in Eq. \eqref{qram} in which qRAM address is provided to the memory access function.
% and the corresponding data is received as output. 
\begin{equation}
 \label{qram} 
 \sum_i\alpha_{i}\ket{i}_{addr}\ket{0}_{data}\longrightarrow \sum_{i}\alpha_{i}\ket{i}_{addr}\ket{\psi}_{data} 
\end{equation}
where, $\alpha_{i}\ket{i}_{addr}$ is the address for accessing the qRAM and $\ket{\psi}_{data} $ is the memory data update.
% \begin{figure}[h]
%   \centering
%   \includegraphics[width=\linewidth]{Quantum_Logic_Gates.png}
%   \caption{Quantum Logic gates}
%   \description{}
%   \label{fig:Qgates}
% \end{figure} 

\subsection{Quantum bolts:  A bit of linear algebra }
Linear Algebra (LA) is a branch of mathematics, to study vector spaces (complex or real) and the operations defined over them. It forms the basis of the QM as well as the QC. In this subsection, minimal but exhaustive components of the LA will be described as the mathematical foundation of the QC. The detailed discussions are available in the textbook \cite{Nielsen2002, Djordjevic2022, lay2003linear}. The LA is used to represent the qubit states, logic gates, their linear operations, and the output of the application of the quantum circuits. The understanding of quantum programs will require the understanding of the following key terms of LA which are described as follows:

    \paragraph{Hilbert Space} A Hilbert space $\mathbb{H}$, is a vector space over a field $\mathbb{F}$ of complex numbers $\mathbb{C}$, defined using inner product $<p.q>$. It must be complete in terms of the norm as defined by $||x|| = \sqrt{<x,x>}$. The QC uses finite-dimensional Hilbert space such as a qubit is defined using two-dimensional Hilbert space. 
    
    \paragraph{Basis vector} A basis vector $\mathbb{B}_{H}$, is defined over a Hilbert space $\mathbb{H}$ as a collection of linearly independent vectors which can span the vector space. Any vector in a vector space can be written as the linear combination of the basic state.
    The number of elements in the basis vector is equal to the dimension of the vector space. A qubit is defined over basic vector $\ket{0},\ket{1}$ and measured over basic vector  $\ket{0},\ket{1}$ or $\ket{+},\ket{-}$.

    % $\{\ket{v}\}_{v\in V}$ such that it maps to 1 for $\ket{v}_{j = i}$ and  0 for $\ket{v}_{j\neq i}$ for every $v\in V$. 
    
    \paragraph{Tensor Product} The tensor product of two vector spaces, $\psi$ and $\phi$ of dimension m and n respectively, is other vector space  $P = \phi \otimes \psi $, vector space of dimension $|P| = mn$. A tensor product is used to define a quantum register.  
    \paragraph{Unitary Operator} A unitary operator $\mathbf{U}$ is a linear operator with a unique property where the inverse can be calculated by its adjoint. The condition can be mathematically described as $U^{\dagger} = U^{-1} $. Every quantum logic gate is defined as  a unitary operator.
    % as can be seen in Fig. \ref{fig:Qgates}.
    \paragraph{Hermitian Operator} Hermitian Operator $\Theta$ is defined as an operator if and only if it has real eigenvalues. The condition can be mathematically described as $\Theta^{\dagger} = \Theta$. All observable (measurable quantities) in QM and QC are hermitian operators. 
    % A hermitian operator is used to implement the measurement gate in a QC. \\
After looking at the QC components and LA entities, next see how these components help in architect a QC and realize it as hardware.
%% Add a table 
\begin{table}[t]
\caption{Linear Algebraic Entities and their application in Quantum computing} 
 \label{tab:LAA}%
\centering
\begin{tabular}{m{2.5cm}m{5.2cm}}
\hline
LA Principle & Quantum Computing Application   \\
\hline
Hilbert Space& A qubit is represented as a unit vector in a 2D Hilbert space over basis state $\ket{0}$ and $\ket{1}$. \\\hline
Tensor Product& Multi-qubit state are represented by a unit vector in the (tensor) product space which in turn defines a quantum register.\\ \hline
Basis Vector& The basis vector of a Hilbert Space or product space represents a definite state of the qubit.\\ \hline
Unitary Operator & Unitary operation on qubits used to represent a quantum operation or logic gate.\\ \hline
Hermitian Operator & A qubit state is measured using a Hermitian operator \textit{i.e.} qubits to classical bits conversion. It projects a quantum state to a basis vector. \\
\hline
\end{tabular}
\end{table}

\subsection{Quantum Computer Architecture} 
\begin{figure}
    \centering
    \includegraphics[scale=0.22]{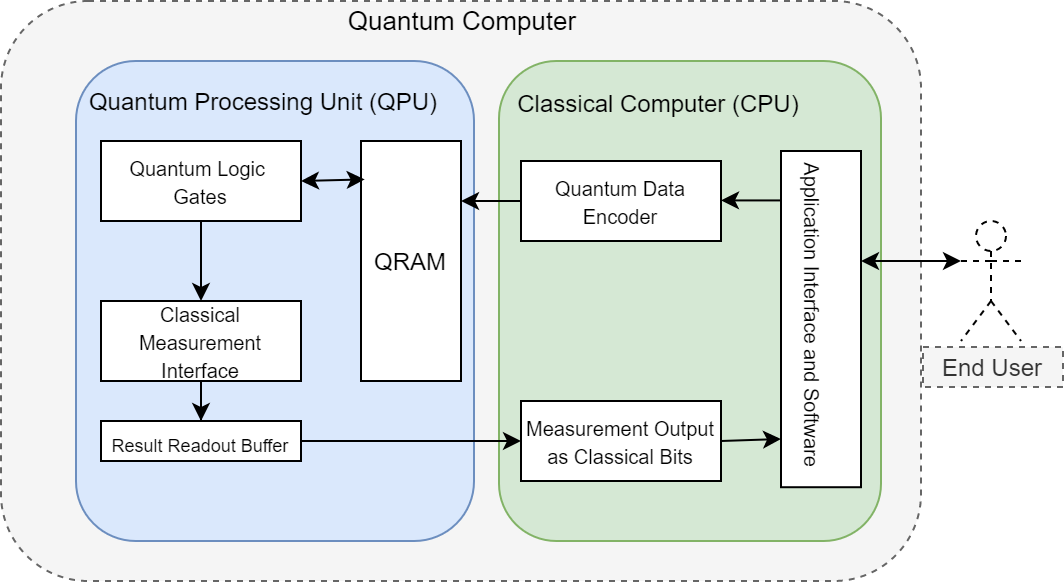}
    \caption{Quantum Computer: A Quantum computer is shown in the descriptive diagram which shows QPU as a quantum process and a classical computer is used to encode input data and receives output as classical data.}
    \label{fig:QCArch}
\end{figure}

A QC hardware is computing devices that are the physical realization of QC via complex electronic circuitry \cite{ladd2010quantum}. It requires implementation of qubit, quantum gate and the read write interface.
% They are more close to nature as they exploit the continuous nature instead digital nature of quantum computers. 
% The primary QC component that distinguishes it from CC is qubits and quantum gates. 
The most challenging part of developing a QC is the implementation of qubits without decoherence. The QC can work as a computation accelerator for CC as shown in Fig. \ref{fig:QCArch}, where a QC is shown as an ensemble of CC components and quantum processor. \cite{qcaciliberto2018quantum}. \\ 

\subsubsection{Quantum Development Technologies} Different methods \cite{ladd2010quantum} have been evolved for the implementation of a physical qubits. These technologies use different chemical and physical properties of the material and engineering technologies for the implementation of QC \cite{Hidary2021} as described below:
\begin{itemize}
    \item Neutral Atoms: Neutral Atoms (in order of millions) are cooled at very low temperatures and controlled with the help of a laser to form a magneto-optical trap. These trapped neutral atoms are then realized as qubits \cite{henriet2020quantum}.
    \item Nuclear Magnetic Resonance: It provides spin qubits based on the magnetization of atoms. 
    \item Nitrogen-Vacancy center in diamond: A nitrogen atom is replaced with a carbon atom in a diamond to create a vacancy center in a diamond. The paramagnetic defect can be treated as a qubit. 
    \item Photonic Qubit: Photonic qubit utilizes the photons and optical components for the implementation of qubits.  
    \item Silicon-Spin Qubits: It utilizes the silicon-based implementation of the spin qubits \cite{maurand2016cmos}.   
    \item Superconducting Qubits: It creates a qubit with the help of the cooper pair and Josephson junction. The SC qubits requires very low temperatures (in order of mK) and microwave control signals \cite{kjaergaard2020superconducting}.
    \item Topological Quantum Computation: It utilizes anyons for creating a coherent QC \cite{nayak2008non}. 
    \item Trapped Ion (TI): It utilizes the ionized atom controlled by magnetic fields to realize linear qubits \cite{bruzewicz2019trapped} .
\end{itemize}
These technologies are utilized by different companies to produce a variety of QC. The SC qubits and TI qubits are mostly used in developing a QC. The devices based on these technologies require a very peculiar environment to operate, such as cooling requirement and environment isolation. So, to fabricate and orchestrate a QC based on these qubit implementation, requires very precise and an intensive engineering. To provide a guideline of QC hardware development, standard QC architecture will be required which is described in the next subsection. 

\subsubsection{Five layer Architecture of a Quantum computer} 
\begin{figure}[ht]
    \centering
        \includegraphics[scale=0.25]{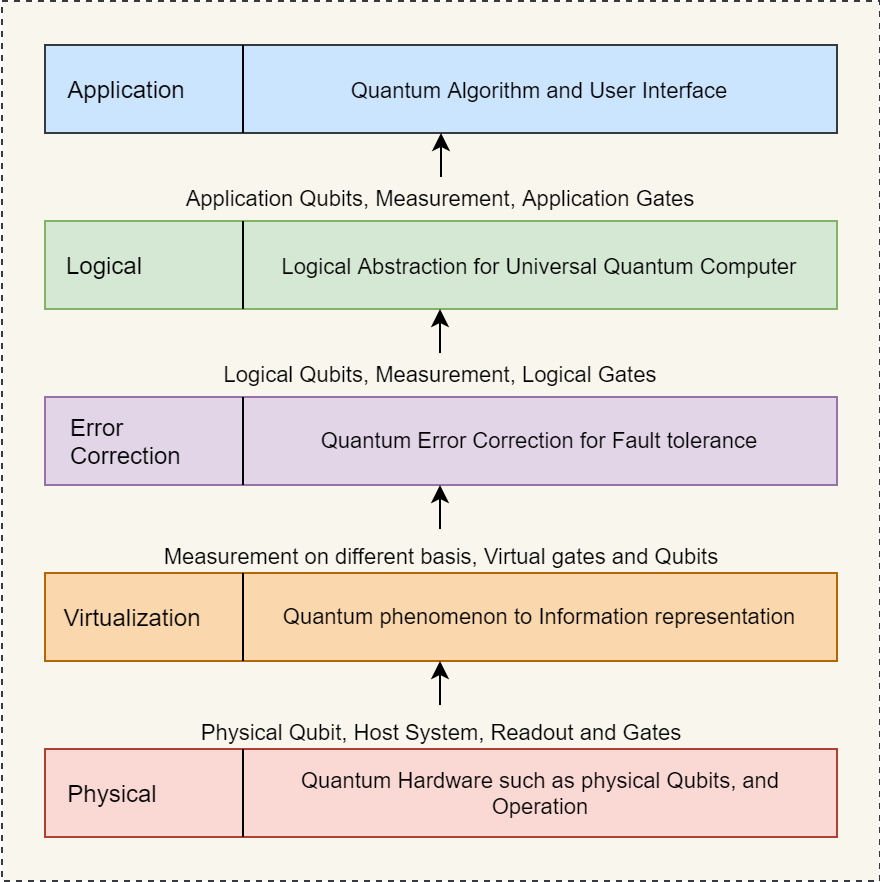}
    \caption{Five Layered Architecture of Quantum computer consisting of Application, Logical, Error Correction, Virtualization, and Physical layer and interface between each layer.}
    \label{fig:layerqc}
\end{figure}

The five-layered architecture \cite{layeredQCPhysRevX.2.031007} of a QC is shown in Fig. \ref{fig:layerqc} which consists of physical, virtual, error correction, logical, and application layer described as follows:

    \paragraph{Physical Layer} The physical layer consists of physical qubits, related hardware, and their enclosed environment. The quantum gates are realized using different physical operations. The physical layer may be fine-tuned as per different qubit technology such as microwave irradiation \cite{kjaergaard2020superconducting}  or laser operations \cite{wang2020integrated}. The environmental interaction while reading or manipulating a qubit causes the occurrence of noise. Noise deteriorates the quantum state which needs to be managed properly. The physical layer is interfaced with the virtual layer that provides physical qubit, readouts and other gates.
    
    \paragraph{Virtual Layer} In the virtual layer the quantum states are converted into information primitives such as qubit and quantum register. It also converts physical operations to logical gates. It also helps in creating a coherent system. It provides virtual qubits, logic gates and measurements on a different basis to the error correction layer.

    \paragraph{Error Correction Layer} The error correction layer is used to handle errors occurred due to noise. A fault tolerant QC will be required for high capacity QC. The Quantum Error Correction (QEC) \cite{knill2000theory} technology is deployed to correct the error incurred. The QEC pumps the entropy out of the system in the form of error syndrome and creates logical qubits and logical gates which are fed into the logical layer.
    \paragraph{ Logical Layer} The logical layer creates an abstraction required for the universal quantum computer \cite{deutsch1985quantum}. It uses the fault-tolerant structures of logical qubits and logical gates and processes them for creating arbitrary gates to be used by the application layer. 
    \paragraph{ Application Layer} The application layer is the interface where user interacts with the system by providing the quantum encoded \cite{dataencoding} input data. Quantum algorithms are executed in the application layer. The application layers are free from underlying layers as it works on application qubits and arbitrary logic gates. 

The above five-layered architecture creates a basis for the development of scalable fault-tolerant quantum computers. After designing a QC the performance metrics needs to be defined for evaluating its efficiency. \\ 
\subsubsection{Performance Metrics of a Quantum Computer}
The performance metrics provide a quantitative measurement of QC performance. The performance of the QC depends on the number of qubits, qubit-state coherence, number of operations per unit time, and the circuit depth \cite{Resch21Bench,wang22bench}. IBM defines three key metrics \cite{wack2021quality} for measuring QC performance as Scale, Quantum Volume (QV), and Circuit Layer Operations Per Second (CLOPS). Table \ref{tab:kpm} illustrates the key performance metrics of the QC. Scale simply represents the system's capacity as the number of qubits. QV represents the size of the largest square circuit of random two-qubit gates that can be successfully executed on the given quantum processor. CLOPS is  a representation of the speed \textit{i.e.}, how much time it will take to execute a quantum circuit. 

\begin{table}[ht]
\caption{Key Performance Metrics of a Quantum Computer} 
 \label{tab:kpm}%
\centering
\begin{tabular}{m{2.5cm}m{2.5cm}m{2.5cm}}
\hline
Scale & Quality & Speed   \\
\hline
Measured with the number of qubits.& Measured in the terms of QV. & Measured by the time taken for executing a circuit.\\
Expresses how much information that can be encoded and processed. & Defined as the size of the largest square circuit with two-qubit gates that can be executed.& Execution time is measured as CLOPS. Higher CLOPS implies higher speed.\\
Requires high coherence, low cost, high reliability. & Requires low operation error rate. & Requires seamless. synchronization among classical and quantum devices.\\
\hline
\end{tabular}
\end{table}
After looking at the QC fundamentals, design, architecture and performance metrics we are ready to look at QC based solutions. So, quantum algorithms are introduced in the next section.
% \begin{figure}[h]
%   \centering
%   \includegraphics[ width=\linewidth]{perform1.jpg}
%   \caption{Performance Metric of Quantum computers}
%   \label{fig:PerformanceMetric}
% \end{figure}

\section{Blueprint: Quantum Algorithms}
\label{sec:3}
 The quantum advantage is based on the assumption that QC can solve valuable problems that present classical supercomputers cannot solve \cite{aaronson2016complexity}. The QS can be verified by executing efficient quantum algorithms on QC, for which no better solutions exist in the CC. 
 
 % Algorithms are building blocks for end-to-end quantum solutions, whether it's related to quantum simulation or QML. 
  A quantum algorithm is a step-by-step procedure to solve a problem using QC methods where each step can be executed on a QC/simulator. The input data should be encoded as qubits and output data is classical bits. To implement an algorithm QC supports two different quantum paradigms as discussed in the next subsection. 
%% comment classification of Quantum algorithms can be given here
  
\subsection{Quantum Computing Paradigm} There are several QC paradigm such as Universal gate model, adiabatic model, topological computing, continuous-variable model and quantum annealing model. Out of which two prominent computing models in QC are the universal gate model and  the quantum annealing\footnote{Adiabatic model is more generalized model of quantum annealing, but it is less successful in practical applications.}.  
    \subsubsection{Universal Gate Model} Universal gate model of QC is analogous to the gate model of CC, where computational operations are implemented by universal quantum gates. The interactions between qubits are implemented as quantum gates, and the interaction affects the qubit state.  
    \subsubsection{Quantum Annealing \cite{finnila1994quantum}} It is a quantum-based meta-heuristic approach to solve large-scale combinatorial optimization problems \cite{dwave}. It can solve problems expressed as an optimization problem. It is applicable to two types of problems.
    \paragraph{Optimization Problem} Annealing-based optimization translates the objective function as the energy function (Hamiltonian) of the system \cite{hen2016quantum}. The lowest energy, \textit{i.e.} ground state energy is approximated as the minimum value of the objective function or solution of the optimization problem. 
    \paragraph{Sampling Problem} Annealing sampling problem \cite{yulianti2022implementation} is basically a probabilistic sampling problem where sampling is done over many low-energy states of a system. By sampling low-energy states an energy landscape is created that can provide a model of reality. A particular sample gives a system state with certain instances of parameters that can be utilized in model accuracy.  
% Quantum annealing uses quantum physics to find low-energy states of a problem and, therefore, the optimal or near-optimal combination of elements.
% \begin{itemize}
%     \textbf{Optimization Problem}: In an optimization problem, the energy of the system is mapped with the problem and the ground state energy of the Hamiltonian is approximated as the minimization of the function. Quantum annealing simply uses quantum physics to find low-energy states of a problem and therefore the optimal or near-optimal combination of elements.
%     \item \textbf{Sampling Problem}: Sampling from many low-energy states and characterizing the shape of the energy landscape is useful for machine learning problems where you want to build a probabilistic model of reality. The samples give you information about the model state for a given set of parameters, which can then be used to improve the model.
% \end{itemize}
\subsection{Complexity Classes in QC}
\begin{figure}[ht]
  \centering
  \includegraphics[width=\linewidth]{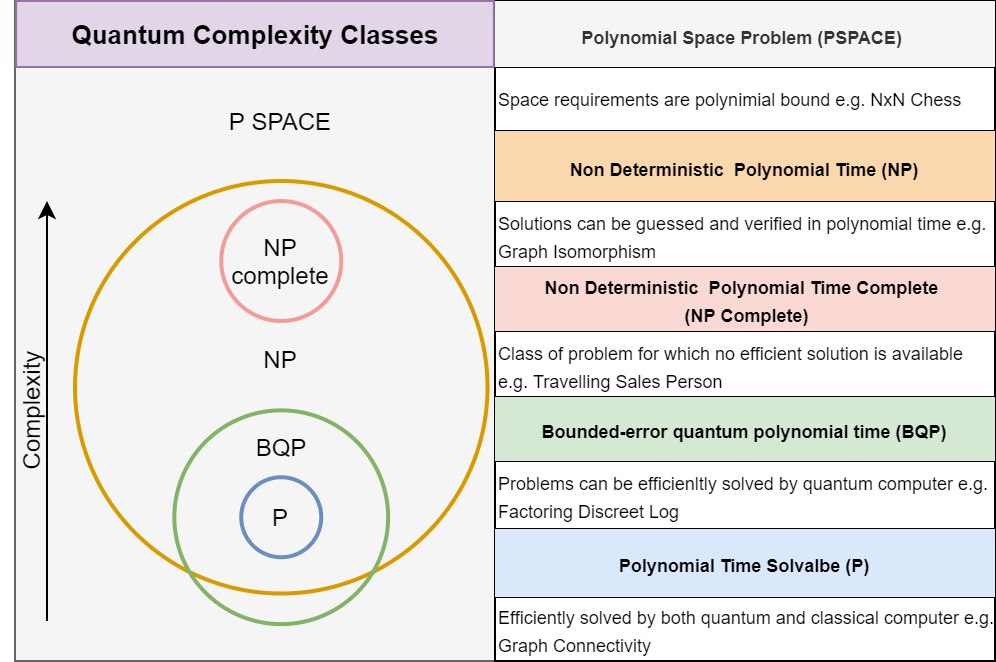}
  \caption{Complexity classes and their description.}
  \label{fig:Complexity}
\end{figure}
 
 The complexity classes are designed to classify algorithms based on their resource requirement such as execution time and run-time space in the worst case scenarios. To understand the power of a QC, complexity analysis is required to understand which classes of problems are efficiently solvable by QC. The problems are grouped based on their complexity to form a complexity class as provided in \cite{cormen2022introduction}, containing P, NP, and NP-complete complexity classes. With the evolution of QC it is appended with new classes that can efficiently be solved by QC but not by CC \cite{bernstein1993quantum}. The quantum complexity classes \cite{Hidary2021}\cite{complexityaaronson2008limits} are summarized in Fig. \ref{fig:Complexity} and described \footnote{Here P, NP, and NP-complete class description is not provided with that can be extracted from a standard textbook.} as shown below :
   \paragraph{PSPACE} The Polynomial SPACE (PSPACE), represents a class of problems where the space requirement is polynomially bounded in the input size.   
   % \item BPP: Bounded Error probabilistic polynomial time is a class of problem for which a randomized polynomial time solution exist which 75\% success probability.
    \paragraph{BQP} The Bounded-error Quantum Polynomial (BQP) time is a class of problems that can give a correct result with probability in polynomial time.  
    \paragraph{EQP} The Exact Quantum Polynomial (EQP) time \footnote{It is a subset of BQP class but for simplicity, it is not shown in Fig. \ref{fig:Complexity}.} is a BQP problem where the success probability is 100\%.  
     
After looking into the complexity the standard quantum algorithm is discussed along with the classification.
\subsection{Standard Quantum Algorithm}
Quantum algorithms are generally realized by a quantum circuit that acts on input qubits and outputs the classical bits. It can be executed on a QC, simulated on a CC, or over a synchronized orchestration of CC and QC. Quantum algorithms can be classified into different categories as shown in Fig. \ref{fig:Algoc} and described as follows:
 \begin{figure}[ht]
        \centering
        \includegraphics[ scale=0.4]{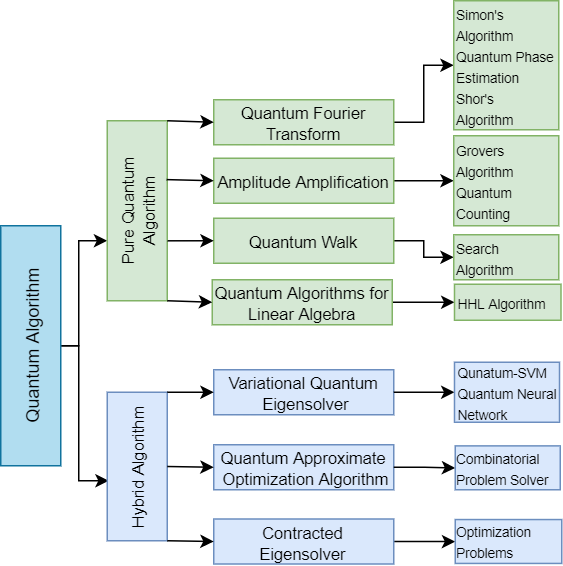}
        \caption{Quantum algorithm classification: Quantum algorithms are classified into pure and hybrid quantum algorithms with further sub-classes and example algorithms. }
        \label{fig:Algoc}
        \end{figure}

    \subsubsection{Pure Quantum Algorithm}
    Pure quantum algorithms are designed to run entirely on a QC. They do not utilize the CC for computation but for input/data. It can be further grouped based on the core approach/paradigm listed below.

        \paragraph{Quantum Fourier Transform (QFT)} QFT \cite{QFT2006moore} is a linear transformation of qubits that plays a role in data encoding by converting the input data coefficients to its Fourier transform. It is one of the core components for many quantum algorithms as its classical counterpart, Fourier transformation, which converts a signal in time domain to the frequency domain. QFT is utilized in Shor's algorithm, discrete log, and quantum phase estimation algorithm. QFT transform a quantum state as $\ket{X}$ to a transformed state $\ket{Y}$ by using the equation Eq. \eqref{eq:qft}
        
        \begin{equation}
        \label{eq:qft}
        \ket{Y} = QFT(\ket{X})= QFT(\sum_{j=0}^{N-1}x_i\ket{i})
        = \sum_{j=0}^{N-1}y_k\ket{k},    
        \end{equation}
        where, $ y_k = \frac{1}{\sqrt{N}}\sum_{j=0}^{N-1}x_j\omega_N^{jk}$ and $\omega_N^{jk} = e^{2\pi i \frac{jk}{N}}$. \\

         \paragraph{Amplitude Amplification (AA)} It  is a class of algorithms that generalizes the Grover search \cite{grover1997quantum} on an unstructured dataset \cite{brassard2002quantum}. Grover's search algorithm is based on dividing the search space into  good and bad subspaces and attempts to map any given quantum state to the good subspace. Grover search algorithm creates a superposition of all the states in the starting, whereas, in the generalized version, it considers the initial bias. In the AA algorithm, if good states are known, they can be incorporated while creating the superposition of all the states. The AA algorithm uses unitary operation such as Grover iteration \cite{kwon2021quantum} to increase the probability of the desired solution.
        \paragraph{Quantum Walk} Quantum walk \cite{venegas2012quantum} is the quantum counterpart of the classical Markov chain \cite{magniez2007search}, which provides the search result of the marked element in a graph after a complete walk. With the help of superposition, all paths can be traversed simultaneously in a quantum walk, providing a quadratic speedup to its classical counterpart. Interference also helps speed up by cancelling out the wrong path traversed. There are two variants of the quantum walk; A coined quantum walk covers the vertices of the graph, while the Szegedy quantum walk \cite{Szegedy2004Quantumwalk} covers the edges of a graph.
        \paragraph{Quantum algorithm for linear Algebra} Quantum algorithms are also helpful in solving the system of linear equations, primarily a matrix inversion problem. Even though matrix manipulation is a polynomial time algorithm it becomes complex for large-size matrices. The HHL(Harvard-Howarth-Low) \cite{harrow2009quantum} algorithm has been proposed with complexity as $\mathcal{ O } (\log(N)s^{2}\kappa^{2}/\epsilon)$, which is a very significant improvement for machine learning and data analytics. The HHL is a quantum version of classical algorithms for solving linear systems, such as Gaussian elimination and the conjugate gradient method.
  
\subsubsection{Hybrid-Classical Quantum Algorithm}
\begin{figure}[hb]
  \centering
  \includegraphics[ scale=0.25]{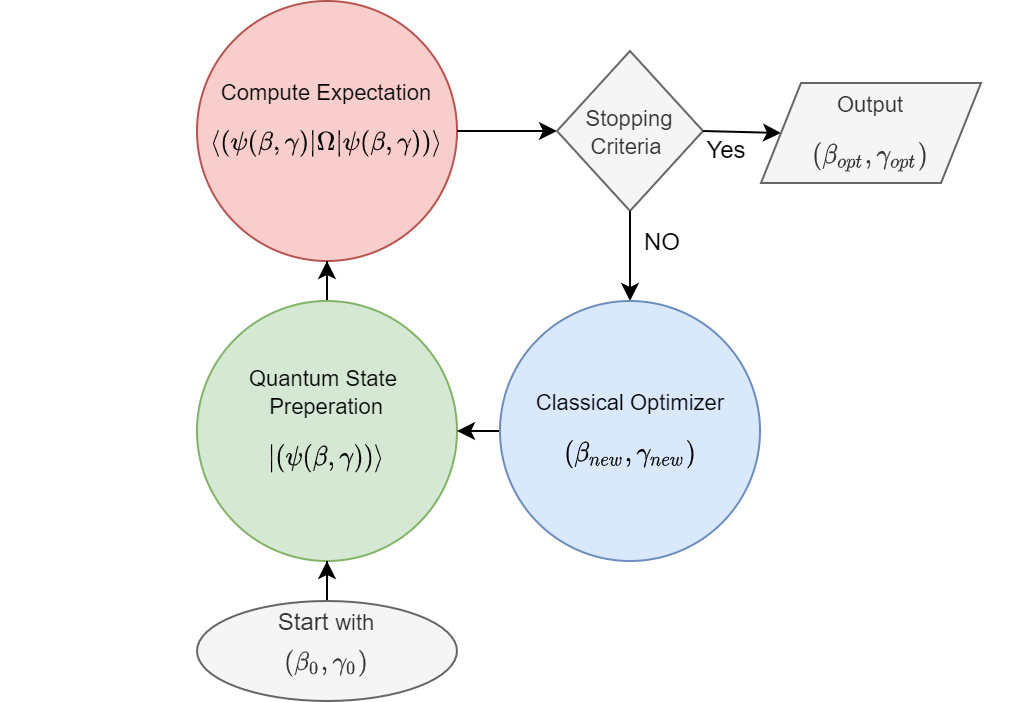}
  \caption{Quantum Approximate Optimization Algorithm (QAOA) schematic diagram: In the Figure, the QAOA algorithm is described where the classical optimizer is used along with quantum-based expectation evaluation.}
  \label{fig:QAOA}
\end{figure}

    The hybrid algorithms are designed to utilize the combined power of QC and CC. In a hybrid algorithm, some portion of the algorithm runs on the CC and rest on the QC. They are useful in NISQ-era devices since the quantum memory has high decoherence and intermediate results can be stored on a CC. They cover QML solutions, annealing-based problems as well as physical system energy minimization problems. The following approaches are present for  hybrid classical/quantum algorithm as discussed below:  
 \begin{figure}[t]
  \centering
  \includegraphics[ scale = 0.42]{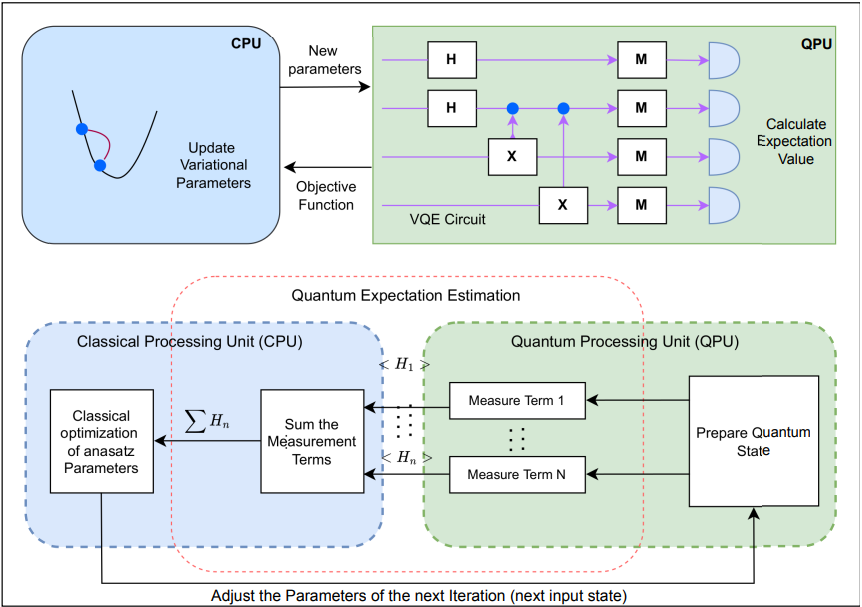}
  \caption{Variational Quantum Computing: The given figure depicts a hybrid algorithm where the classical optimizer is optimizing the ansatz parameter and quantum computers are used to calculate the expectations.}
  \label{fig:VQE}
\end{figure}
    \paragraph{Variational Quantum Eigensolver} In NISQ-era, the Variational Quantum Eigensolver or Algorithm (VQE/VQA) has emerged as the most useful class of quantum algorithms \cite{tilly2022variational}. VQE/VQA is the base of all QML and DL algorithms as well as for optimization problems \cite{zeguendry2023quantum} . It takes a classical physics phenomenon of finding the ground state energy of a system. VQE works on mapping the state energy of a molecule to loss function of QML or the objective function of an optimization problem. Then it uses a variational method to find the ground state energy that represents the solution, \textit{i.e.,} optimized value or minima of the loss function. The VQE utilizes the classical optimizer as shown in Fig. \ref{fig:VQE}. It requires an ansatz $\ket{\psi(\vec\theta)}$ which a trial state with a parameter $\theta$ and the variational method try to find the optimal value of the $\theta$  as shown in Eq. \eqref{eqn:vqe}
        \begin{equation}
        \label{eqn:vqe}
        \min_{\vec\theta} \langle\psi(\vec\theta)|\Omega|\psi(\vec\theta)\rangle,
        \end{equation}
        where, $\Omega$ is the Hamiltonian operator and $\theta$ is the optimization parameter.     
    \paragraph{Quantum Approximate Optimization Algorithm (QAOA)} It is used to solve a combinatorial optimization problem to provide an approximate solution by utilizing VQC layers of quantum evolution \cite{herrman2022multi}. It is  specially designed variational method to solve problems such as Max-Cut and Min-Cut. It starts with initial state preparation unitary $U({\beta}, {\gamma})$ where 
        $\beta$, $\gamma$ parameters of the form as given in the equation Eq. \eqref{eq:qaoa} 
        \begin{equation}
            \label{eq:qaoa}
            U(\beta) = e^{-i \beta \Omega_{m}}, 
            U(\gamma) = e^{-i \gamma \Omega_{p}}, 
            \end{equation}
            where, $\Omega_{p}$  is problem Hamiltonian and $\Omega_{m}$ is the mixer Hamiltonian. 
        The goal of QAOA algorithm is to find the $(\beta_{opt}, \gamma_{opt})$ so that the quantum state $\ket{ \psi(\beta_{opt}, \gamma_{opt})}$
        encodes the solution to the problem. Fig. \ref{fig:QAOA} provides the schematic diagram of the QAOA \cite{QAOA} algorithm.  

After establishing the QC foundation, architectural components and algorithmic design principles, let's move ahead towards the implementation methods. So, in the next section, we introduce quantum development platforms including quantum simulators and cloud-based QC in a way that new users can easily understand, identify and pick a correct development platform. 
% \begin{table}[ht]
% \caption{Summary of quantum computing algorithms}
%  \label{tab2}%
% \centering
% \begin{tabular}{m{2.8cm}m{1.2cm}m{3.5cm}}
% \hline
% Quantum Algorithm &  Type & Objective  \\
% \hline
% % \multirow{}{}{}
% Deutsch–Jozsa algorithm\cite{deutsch1992rapid}   &  QFT & Problems requiring exponential queries\\
% Bernstein–Vazirani algorithm \cite{bennett1997strengths}  & -& Efficient solution of black-box problem\\
% Simon's algorithm \cite{brassard1997exact}  &-& Faster computation, Speedup\\
% Shor's algorithm \cite{shor1994algorithms} &QFT& Integer factorization and discrete algorithm problems\\
% Grover's algorithm \cite{grover1997quantum} & GO & Searching unstructured database for marked entry\\
% Quantum counting & -& Generalised search\\
% Quantum approximate optimization algorithm & Hybrid & Solution of graph theory problems\\

% \hline
% \end{tabular}
% \end{table}

\section{Sack of Tools: Quantum Computer Programming Toolkit}
\label{sec:4}
After exploring the QC programming paradigm and standard algorithm, the quantum developments tools are introduced that will be used in programming a QC or quantum simulations. The QC evolution in recent years provides a handful of quantum development tools, including standalone simulators, software development kits (SDK)'s and programmable QC. These tools are majorly free, open-source and publicly available \cite{sdklist} to academia, industry, and naive programmer. These tools are different in terms of approach, target applications, or underlying platforms. So, a deep understanding with clear idea of these tools will be required for effectively utilization of expressive power of quantum algorithms with ease. 

% different tool carries a unique set of features but may miss others. So, to use these tools, a deep dive is required to get a clear idea of them and a relative comparison among them. 

\subsection{Quantum Development Tools}
Quantum algorithms can be tested on a quantum simulator or actual QC. Both options have some advantages and limitations. With recent hardware developments, the QC are now available with a capacity of (5 to 100 qubits) on SC qubits and up to 5000 qubits as a quantum annealer. The quantum hardware is available via QCC service providers such as Amazon Braket and IBM Qiskit etc. QCC also provides a development platform that is really helpful for hardware access and algorithm execution. Quantum simulators are available as standalone quantum software as well as a service over the cloud. 
% Quantum Simulators are computer programs written in high-level languages providing support for qubit implementation, unitary operations, and quantum operations over a CC. 

Quantum simulators are designed to mimic QC and their state evolution via the CC. It helpful in testing the algorithm before executing over real hardware due to the limited availability and computation cost of a QC. Also the NISQ era devices are noisy which hinder the actual performance of the algorithms. So, quantum simulations are more suitable for design and development of quantum algorithm since simulation provide noise free environments and can explore the full potential of quantum algorithms. Even though quantum simulations are helpful, they have the practical limitation of main memory requirements. The top supercomputer simulation for full wave quantum simulation has reached only up to 48 qubits \cite{park2022snuqs}. So a high capacity quantum hardware will be required for algorithm relying with high number of qubits. Simulation can also help in developing solutions for NISQ-era devices by incorporating noise models in circuit executions. Different companies provide an integrated development platform to access a variety of quantum hardware and simulators.  
% With the current computer, a 32 qubits system can be simulated with ease but beyond 40 qubits is not feasible with current RAM technology. 
Considering these aspects the quantum development platforms can be categorized in the following manner:
    \subsubsection{Quantum Simulators} Quantum simulation provides advantages to test quantum algorithms over the existing classical hardware limited to several qubits. It can be defined as a system that actively uses quantum effects to solve a problem related to  model systems. 
    % Through them, real systems and also revealing information about an abstract mathematical function relating to a physical model. 
    
    % There is a long list of quantum simulators that may attempt to utilize the expressive power and developer preference of different languages. In our discussion only relevant simulators will be discussed, not all of them. \\
    The main memory requirement of quantum simulation exponentially increases with the number of qubits as simulating a qubit system, two complex numbers are required for probability amplitudes for basis $\ket{0}$ and $\ket{1}$. For two qubit four complex numbers will be required per basis states ($\ket{00}\ket{01}\ket{10}\ket{11}$). Likewise, for the N qubit system, total $2^N$ basis states will be required. So. for simulating 32 qubits, $2^{32}$ storage bits are required (\textit{i.e.,} order of 4GB) and 64 qubits, $2^{64}$ storage bits are required (\textit{i.e.}, order 1000 PB). Simulating such a high memory requirements algorithms is not feasible with current hardware, even with utilizing the top supercomputers maximum simulated qubits is 48 qubits. Even though different qubit representations (such as tensor network \cite{orus2014practical} or density matrix \cite{fano1957description}) can be utilized to reduce the main memory requirements. 
    % 1000 PB memory creation is not feasible with current hardware, so simulating QC  will not go beyond 30-40 qubits.
    % As an example, for simulating 32 qubits total size of(complex)$\times2^{32}$ storage bits are required which can be expressed in order of 4GB, but for simulating 64 qubits size of(complex)$\times2^{64}$ bits are, \textit{i.e.}, order 1000 PB which is far beyond the limits of today's supercomputers. So, simulating quantum computing will not go beyond 30-40 qubits.\\
    Based on the qubit representation the quantum simulator can be categorized into three types as listed below:
    \paragraph{State Vector (SV)-simulator} An SV-simulator utilizes the state vector representation \cite{peres1984state} of qubits. SV-simulator works on full wave state representation, and a circuit simulation leads to the sequential application of quantum gates to change the state vector. It stores all the possibilities and its space requirements are exponentially proportional to the number of qubits.
    \paragraph{Density Matrix (DM)-simulator} DM-simulator utilizes the density matrix representation \cite{fano1957density} of qubits. DM-simulation is useful in mixed state quantum representation where matrix elements provides the probability to be in a particular state. It stores the full density matrix of a state, and a circuit simulation leads to the sequential application of quantum gates to change the density matrix.  
    \paragraph{Tensor Network (TN)-simulator} TN-simulator is a quantum simulator that utilizes the graph representation with nodes and edges \cite{cabra2012modern} \cite{orus2019tensor}. Nodes represent the quantum gates or qubits and edges represent wires/connections between gates. The execution of the TN-Simulator is performed in two phases a rehearsal phase and a contraction phase. In the rehearsal phase, it traverses the whole graph to provide ease of measurement, and in the contraction phase, it performs the execution of the circuit. 

    The memory requirement of SV-simulator is highest, and TN-simulator is minimum. The density matrix memory requirement is less than SV-simulator and can be more optimized using sparse state representation. The TN-simulator requires less storage but more compute intensive. 
     
    \subsubsection{Quantum Cloud Computer} Cloud computing with its service models (IaaS, PaaS, SaaS) provides Infrastructure, Platform, and Software as service on as pay as per usage basis \cite{mell2011nistcloud}. It reduces the time and cost of infrastructure setup and provides support for dynamic resource requirements. QC hardware setup requires very specific infrastructure such as (dilution, refrigeration, and isolation from outside noise). Till now, all the development are very delicate, and its technology, such as SC qubits, is in the nascent stage of development. Therefore, the best choice is to utilize the cloud computing paradigm for QC as a service \cite{leymann2020quantum} since in-house QC is not a cost and time-effective choice. IBM provides its devices as cloud service with low capacity devices (5 qubits) free of cost and other devices as (ibmq\textunderscore cityname(27 qubits) to Eagle\_r (127 qubits) as pay per use. Amazon, on the other hand, provides different QC (Dwave, Ionq, Regetti, etc.) on the Braket platform. Other service providers such as Microsoft also provide several QC on its Azure quantum cloud platform. Table \ref{tab:2} provides the details of all significant QCCs and their comparison. 
     % Most of the leading development cloud platforms provide both the facility of the simulator and as well as executing programs on an actual quantum computer.

    \subsubsection{Quantum Software Development Kit (Q-SDK) } Q-SDK is a software tool to execute quantum algorithms on a QC, simulators or emulators. It is an integrated platform with high-level programming API, compilers, simulators, access to QCC, and visualization tools. IBM, Amazon, Microsoft, Regetti, and D-wave is actively developing and improving their quantum development platforms. A detailed comparison of the Q-SDK is provided in Table. \ref{tab:qsdk}. 

 Our focus will be primarily on the Q-SDK as it provides complete support for end-to-end development. A detailed discussion on selected Q-SDK- Cirq, Qiskit, Pennylane, Braket and Microsoft-QDK along with brief discussion on other tools is provided in the rest of the section. Each Q-SDK will be discussed based on its simulation features, noise modelling, hardware access, pricing etc. 
 
\subsection{Cirq} Cirq \cite{Cirq} is an open-source, python-based quantum development framework provided by Google Quantum AI \footnote{\url{https://quantumai.google/}}. It is designed to simulate universal gate quantum computers and also provides the facility to execute a program on the QC hardware. It is suitable for creating, editing, and executing optimized quantum circuits. 
\subsubsection{Cirq Simulation tools}
It supports majority of quantum gates including standard gates and custom gates. Devices which is a pre-built set of gates, is also supported by Cirq. A quantum circuit is a collection of moments which is the vertical slicing of the circuit. In Cirq, the circuits can be created either by iteration or appending circuit components one by one. The Cirq simulator can work only up to 20 qubits. The Cirq supports the following ways for simulation:
\begin{itemize}
    \item Exact Simulation: In the exact simulation, the Cirq assumes noiseless circuits, which gives the exact expected output from the circuit. 
    \item Noisy Simulation: In the noisy simulation, the noise models are appended with the circuit, which affects the output due to the presence of noise. This simulation tries to match the real Nisq-era hardware. 
    \item Parameter Sweeps: In this simulation, a parameterized circuit is used to obtain the optimal parameter by measuring and changing parameters.
    \item State Histogram: This is used for visualizing and analyzing the output by providing a histogram view of the output.
\end{itemize}

\paragraph{Noise Modelling in Cirq} It provides three ways for appending noise as channels, noise models, and Cirq measurement gates. In noise modelling as channels, both coherent and incoherent errors can be implemented. Coherent errors are implemented as gates, whereas incoherent errors are modelled as some probabilistic operations. Some standard noise models are pre-built in Cirq and custom noise models can also be created in Cirq. The inbuilt noise models such as constant noise, and inverted noise models are available in Cirq. Constant noise models are used to insert simple noise into each gate in the circuit. Whereas the insertion noise inserts noise only at the specified points in the circuits. Measurement parameters provide the modelling for noise incurred while applying the classical measurement step. Invert mask and confusion matrix, are measurement noise models.\\

\paragraph{Quantum Simulator (QSim)} QSim is the quantum simulator available with the Cirq framework and it is an SV-simulator. It is written in C++ and exploits the OpenMP API for efficient quantum simulation using multi-threading and vectorized instructions. It can simulate up to 40 qubits with an intel 90-core xenon processor. Google Cloud supports CPU-based simulation, GPU-accelerated simulations as well as multi-node simulations. A paid plan is required for large circuit simulation using Google Cloud Platform (GCP). 

\begin{figure}[t]
    \centering
    \includegraphics[width=\linewidth]{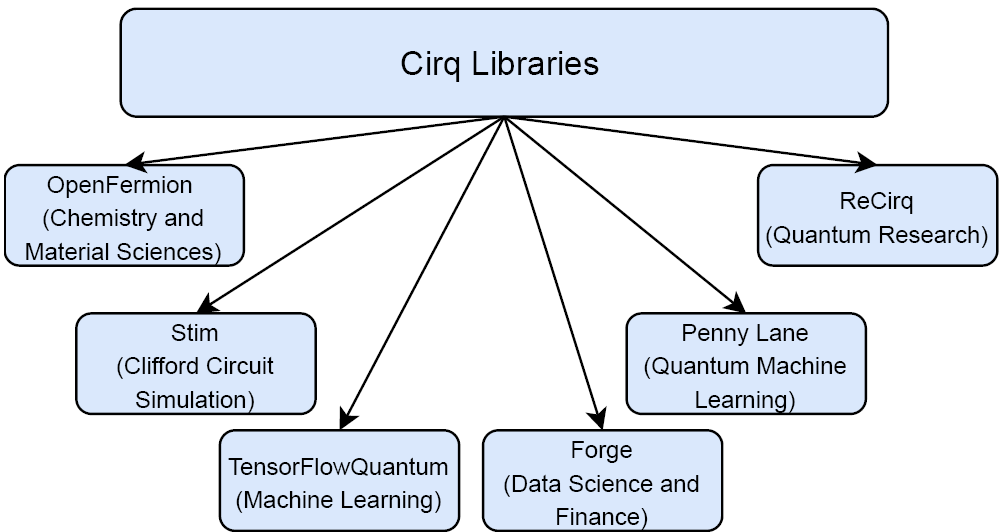}
    \caption{The Ecosystem of Cirq Libraries: Different libraries that can be integrated with Cirq based on the application.}
    \label{fig:cirqeco}
    \end{figure}

A circuit designed in Cirq is a general circuit that can be used on different simulators or hardware. Circuit transformation is used to convert a user-defined circuit to a hardware-specific circuit. Cirq supports an inbuilt transformer to adjust alignment, delaying measurement, which can be added as per requirement. All textbook codes such as Shor's algorithm, VQE, and QFT are directly implementable on Cirq without any additional library.
\paragraph{Cirq Ecosystem} The Cirq ecosystem \cite{cirqlib} supports a vast set of external libraries that can be utilized for application-specific development. 
The following additional libraries the Cirq library (Fig. \ref{fig:cirqeco}) are described as follows:
\begin{itemize}
    \item OpenFermion: It supports quantum experiments for chemistry and material sciences.
    \item TensorFlowQuantum: It is designed for developing QML algorithms by exploiting the power of the existing tensor flow library. It is used to develop the hybrid classical-quantum algorithm.
    \item Stim: It is useful in  Clifford circuit simulation and error correction.
    \item Recirq: It is an extension of the Cirq library with additional cutting-edge quantum methods to extend quantum research. 
    \item Forge: It is proprietary for the specialized domains in data science, finance applications, etc.
    \item Pennylane: It is a library specially designed for the implementation of QML using TensorFlow, PyTorch, and NumPy. 
\end{itemize}

Cirq also supports qudits which is a multilevel quantum system such as qurits (with three levels), ququarts (with four levels), etc. 
\subsubsection{Google QCC}
Google is also extensively working on its QC. It has also developed SC-qubit QC with Foxtail, Bristlecone, and Sycamore in the list. Sycamore processors are now utilized as the core of the Google QCC \footnote{Google Quantum Computing Service, which provides quantum hardware access, is currently not available for public access but it can be accessed by authorized persons.}. The details of all these processors are provided below: 
% \begin{figure}[h]
%   \centering
%   \includegraphics[ width=\linewidth]{syncamore.png}
%   \caption{Google Quantum Processor: Syncamore}
%   \label{fig:syncamore }
% \end{figure}

\begin{itemize}
    \item Foxtail: It is a 22-qubit quantum processor. The qubits are arranged in the 2D unit cells. 
    \item Bristlecone: It was a big leap in quantum computer with a design of 72 qubits quantum computer system arranged in a square grid. 
    \item Sycamore Processor: It is a 54 SC-qubit system with a square grid lattice arrangement of the qubits. It is suitable for NISQ algorithms such as Hartree-Fock (chemistry), QAOA (optimization), and QML.
\end{itemize}
\paragraph{Weber-QC} It is a sycamore processor-based QC, available on the GCP. The Cirq programs can access it to process a quantum algorithm on real hardware but it is available to the permitted users. There are two ways to access the Weber-QC, one is Openswim which is a job queue where the processor is allocated in FIFO order ensuring fair share and the other one is processor reservation which provides hourly reservation to the approved user.  

\begin{figure}[ht]
  \centering
  \includegraphics[ scale=0.24]{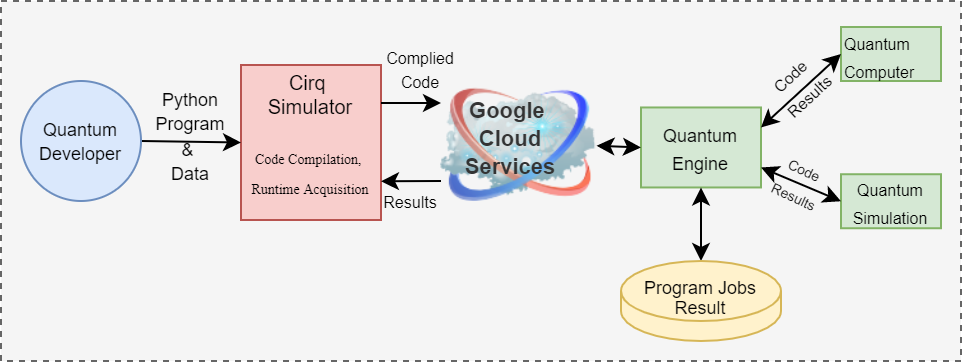}
  \caption{Google cloud computing workflow: A quantum developer interacts with the Cirq library which provides high-level access to quantum hardware and simulator via the quantum engine.}
  \label{fig:QCSworkflow }
\end{figure}

\paragraph{QCC Workflow} The workflow of Google-QCC as shown in Fig. \ref{fig:QCSworkflow } consists of the Cirq framework which interacts with the quantum hardware and simulation software via quantum engine. The quantum programmer can develop different applications using the Cirq framework. The quantum engine does the interaction with the hardware/simulator. It manages the jobs and their scheduling, obtaining the results and providing it to the quantum programmer. 

\paragraph{Quantum Virtual Machine (QVM)} QVM is a way to emulate the QC designed by google. QVM is available with the Cirq framework and it can be obtained by specifying the quantum processor. The output of QVM is similar to an actual QC output so it can be really handy for prototyping, testing, and circuit optimization.

\begin{figure}[ht]
  \centering
  \includegraphics[scale=0.32]{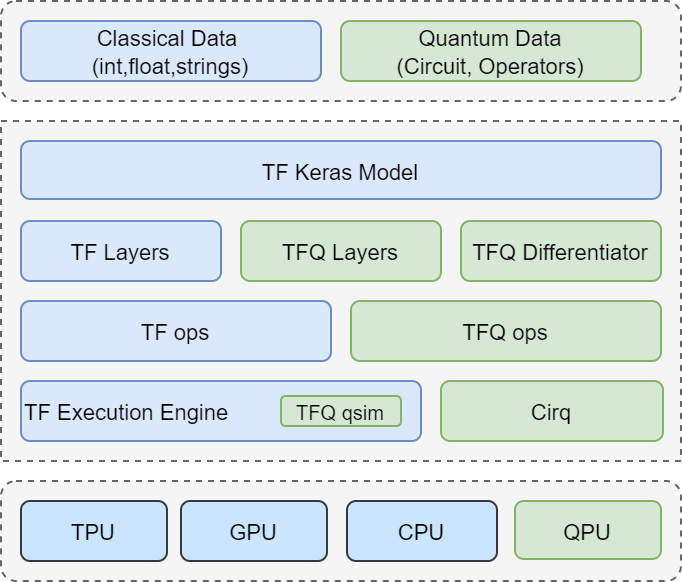}
  \caption{Tensor Flow Quantum (TFQ) Architecture: TFQ architecture shows the integration of classical and hybrid programming modules as well support for executing hybrid algorithms over CPU and QPU.}
  \label{fig:TFQ}
\end{figure}

\paragraph{QML Support} The Cirq framework provides support for developing the hybrid\_classical-QML solutions. Tensor Flow Quantum (TFQ) \cite{broughton2020tensorflow} is a specially designed library for hybrid\_classical-QML. TFQ architecture as shown in Fig. \ref{fig:TFQ}, focuses on quantum-encoded data and building QML-based solutions. It amalgamates QC algorithms developed using Cirq and generates QC primitives compatible with existing TensorFlow APIs. 
% The QML tutorial provides pre-built examples  of Quantum CNN, Quantum Reinforcement learning, MNIST classification, etc. 

\paragraph{Coding interface} The existing Google-Colab platform can be utilized for quantum programming which provides ease of programming and run-time with GPU and TPU-accelerators. 

Limitation of Cirq: The Cirq tool is a pretty handy SDK for QC but it has several limitations discussed as follows:
\begin{itemize}
    \item Limited QC Support: Even though Cirq has the capability to run the same code on a simulator and a QC such as google sycamore processor but these processors are not available for public cloud access. Which makes Cirq primarily a simulation tool. Even though it can be connected to other quantum computer providers through their API.
    \item Community Support: Cirq is having comparatively low community support as compared to other tools.
    \item High Learning Curve: The Cirq is not easy to learn platform due to its inherited complexity, lack of tutorials and limited community which causes a high learning curve for Cirq. 
    \item High End API support: Cirq is designed with low level APIs which makes it complex for a developer to utilize high level API and support for other QCC and simulators.
    \item Memory Constraint: The Google Colab platform provides 3 plans free tier, colab pro, colab pro plus and the corresponding RAM space provided by these plans are 16 GB, 32 GB, and 52GB. With this limited memory, even in the highest-paid plan, the number of qubits will not go beyond 34 qubits.
    \item Visualization Tools: The visualization tools are available but are not as advanced as compared to other tools. It uses ASCII text-based circuit description and Bloch sphere whereas Qiskit provides better visualization such as Qsphere. 
    \item Runtime: It provides run-time units with and colab plus plans which limits the user to access the simulator. even though offline simulation can be done without any additional cost. 
\end{itemize}

% \begin{landscape}
\begin{table*}[ht]
\caption{Quantum Cloud Computers and their comparative Analysis }
\label{tab:2}       % Give a unique label
\begin{tabular}{m{2.2cm}m{1.2cm}m{2cm}m{1.5cm}m{2cm}m{2cm}m{4.3cm}}
\hline\noalign{\smallskip}
Quantum Computer & Architecture &  Availability & Capacity & Topology & Access Type & Key Features \\ \hline
DWave- Advantage &  SC  & Leap, Braket & 5000+ Q & Pegasus  & Free(leap), Paid & Specially Designed for business use cases, 15-way qubit connectivity, the Highest number of qubits, Energy Efficient, Provided end-to-end development stack via Leap \\ 
Google- Weber & SC  &  Google Quantum AI & 54 Q & Square Grid lattice & Paid & It belongs to sycamore processor family, Flip Chip Design, Low error rate \\ 
IBM Q Experience  & SC  & IBM Qiskit & 5-127 Q & Heavy Hex lattice & Free(5-qubit) and Paid & IBM provides a family of a superconducting quantum processor, Multi-Chip stack design, Supports OPENQASM-3, \\ 
IonQ-Processors  & TI & IonQ-Quantum Cloud, Braket, GCS, Azure Quantum & 9-23 AQ  & - & Paid & The family of processors based on Trapped Ion technology, access available on multiple platforms, Freedom to use different development libraries, uses AQ instead of qubits as a metric. less noise supports complex problem  \\ 
Oxford Quantum Circuits & SC &  Amazon Braket & 8 Q& 3D  & Free, Paid & Coaxmon- A 3D chip design to improve coherence in qubits, OQC-Lucy an 8-qubit quantum computer available in Europe only\\ 
QuEra-Aquila\cite{aquila}  & Neutral Atom &  Amazon Braket & 256 Q & Custom &  Free, Paid & It is analog architecture removes the burden of noise, Suitable for simulating quantum dynamics, Custom qubit layout provides more flexible programming, Parallel processing \\ 
Regetti-Aspen-M-2/3 &  SC &  Amazon Braket & 80 Q & Octagonal-3 fold connectivity &  Free, Paid & Designed on the Aspen architecture, scalable multi-chip technology, enhanced readout,\\ 
Xanaudu's QC-Borealis \cite{madsen2022quantum} &  Photonics & Xanadu’s Quantum Cloud, Amazon Braket & 125-216 SSQ & Universal Gate & Free, Paid & First photonic quantum computer with programmable gate, only publically deployed quantum advantage processor, \\ 
% 9 & Pasqal’s QPU & GATE &  & PASQAL & -& -&- & - & - \\
% 9 & Quantum Inspire & Gate & Superconducting Qubits & Qutech & Qutech& 2,5 & Universal gate & cQASM & Free & -\\

% 10 & Pine System & Ion Trap& AQT&  & 20 & -&- & - \\
\noalign{\smallskip}\hline
% \multicolumn{7}{l}{*SC -- SuperConducting, TI -- TrappedIon; **AQ -- Algorithmic Qubits, SSQ -- Squeezed State Qubits;}\\
% \multicolumn{7}{l}{***Free access provides only a limited number of qubits and runtime.}
\end{tabular}
\end{table*}
% \end{landscape}

\subsection{Qiskit}
Qiskit (Quantum Information Science kit) \cite{Qiskit} is a Q-SDK developed by IBM that can be used on IBM quantum lab or as a standalone application. It provides a variety of QC for public access with free and paid plans. 
% It provides both simulations as well programming with real quantum computers.
 Qiskit can be utilized in three ways:   
\begin{itemize}
    \item Circuit library: It provides a complete set of gates as well as pre-defined circuits. Custom gates are also available by using the pulse gates. 
    \item Transpiler: Transpiles high-level Qiskit code to the quantum circuit with the help of basic quantum gates.
    \item QCC access: Qiskit can easily connect to an IBM-QC on which a developed algorithm performance can be verified.
    % It can be utilized to test the code output on real quantum hardware available on the IBM quantum lab platform.
\end{itemize}
\subsubsection{Qiskit Simulation} 
\begin{figure}[ht]
  \centering
  \includegraphics[width=\linewidth]{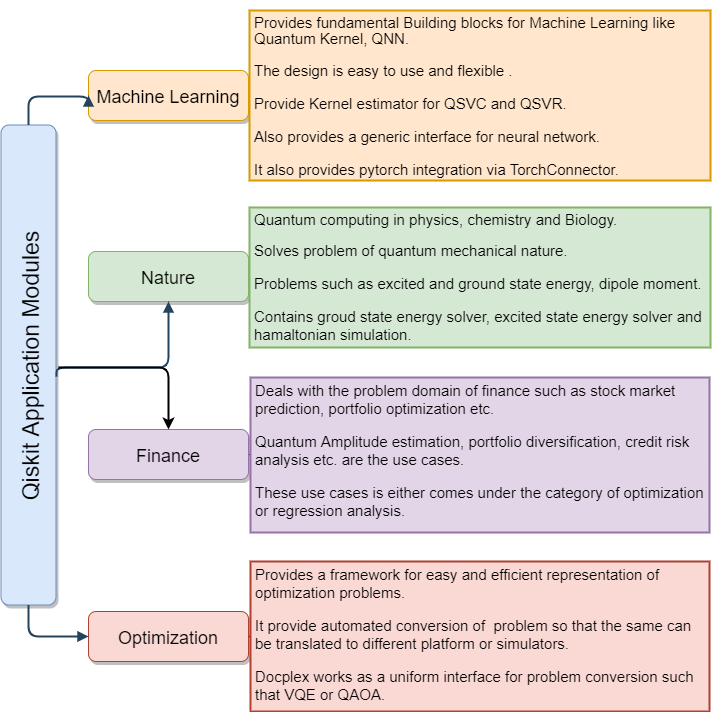}
  \caption{Qiskit application modules: The new Qiskit application modules consist of machine learning, finance, nature, and Optimization modules. The utility of each module is provided in details in their corresponding boxes}
  \label{fig:QM}
\end{figure}

Qiksit provides a variety of high-performance simulation back-ends. It uses a modular architecture with each module supporting different functionality. Let's look at different Qiskit modules to understand its architecture.  
\paragraph{Qiskit Framework} It was initially launched with four modules out of which two are depreciated. The Qiskit modules are described as follows:
\begin{itemize}
    \item Qiskit Terra: It is the core qiskit module, that contains the building blocks for creating and manipulating quantum circuits. 
    \item Qiskit Aer: The Aer module provides a high-performance simulator framework that includes noise modelling in the circuit.
    \item Qiskit Ignis (Depreciated): Ignis module is designed to provide quantum hardware verification, noise characterization, and error correction.
    \item Qiskit Aqua (Depreciated): Aqua is a library for implementing the QML applications. 
\end{itemize}
With the recent update, the  modules are renamed and arranged for targeting different application domains as shown in Fig. \ref{fig:QM} with machine learning, nature, finance and optimization modules. The machine learning module is designed to support QML development, nature is used for quantum physics, chemistry and scientific simulation. The finance module is used for developing quantum applications related to finance such as market prediction and the optimization module specially deals with optimization problems. Three low-level application modules are also available metal, dynamics and experiments. The Qiskit metal module is used for low-level hardware design, dynamics is used for quantum system modelling and Experiments are used for under development that is being developed to support custom experiments over QC. 

\paragraph{Noise Modelling} Qiskit-Aer module is used for noise modelling. It provides the following three classes for noise modelling as described below: 
\begin{itemize}
    \item NoiseModel: It provides a noise model which can be applied to the entire circuit.
    \item QuantumError: It provides completely-positive trace-preserving (CPTP) gate-specific noise models. 
    \item ReadoutError: It provides modelling for classical measurement errors. 
\end{itemize}
\paragraph{Quantum Simulator} It can be obtained from IBM quantum lab API on the run-time. Available simulators in the IBM quantum lab are shown in Fig. \ref{fig:ibmsim}.
\begin{itemize}
    \item ibmq\_qasmsimulator: It is a general-purpose circuit simulator that can be used for any quantum circuit simulation. The noise can also be incorporated during the simulation. It supports a simulation of 32 qubits QC.
    \item simulatorstatevector: It is an SV-simulator that simulates a quantum circuit using the wave representation and operations on it. It supports noise modelling and the capacity is of 32 qubits. 
    \item Simulator\_mps:  It is a TN-simulator that represents a qubit state as a matrix product state. It doesn't support noise modelling. Its capacity is up to 100 qubits. It is more useful in simulating a system with weak entanglement.

\begin{figure}[ht]
  \centering
  \includegraphics[scale=0.26]{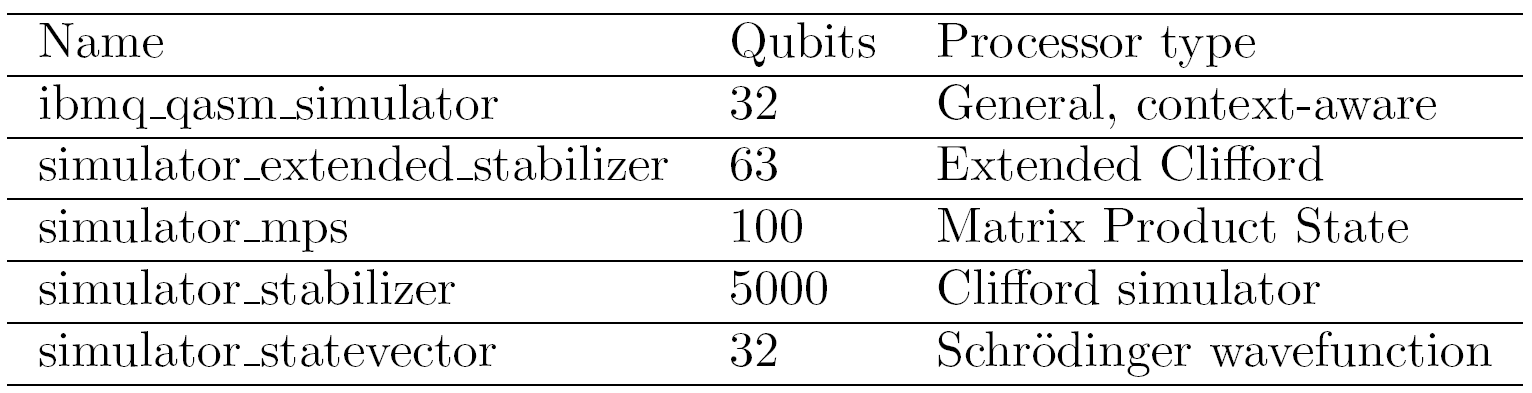}
  \caption{IBM Quantum Simulator: The IBM compute resource page provides a list of compute resources available at that time. The simulators resources have been captured from \url{https://quantum-computing.ibm.com/services/resources}}
  \label{fig:ibmsim}
\end{figure}

% \begin{landscape} 
\begin{table*}[ht]
\caption{Quantum Software Development kit(SDK)/Quantum Simulator and Their Comparative Analysis }
\label{tab:qsdk}       % Give a unique label
\begin{tabular}{m{1.3cm}m{1cm}m{0.8cm}m{1.5cm}m{1.5cm}m{1.2cm}m{1.2cm}m{1cm}m{5cm}}
\hline\noalign{\smallskip}
Platform & Type* & Open-Source & Developer & PL & Standalone-cloud** & Pre-build Algorithm & QML support  & Key Features\\
\hline\noalign{\smallskip}
QuEST  &  S & \checkmark & University of Oxford&  C &S & $\times$ & $\times$  & High-performance simulator, support generalized gates, state vector, and density matrices. distributed, GPU accelerated, QASM support \\ 
\hline\noalign{\smallskip}
QXsimulator &  S & \checkmark & QuTech & cQASM/C++ & S& $\times$ & $\times$  & A large number of qubit (34 qubits), back end Emulation, Noise Modelling\\ 
\hline\noalign{\smallskip}
Quantum++ & S & \checkmark & SoftwareQinc & C++ &S & $\times$ & $\times$ & Multi-threaded quantum simulation library, also support reversible classical logic gates \\ 
\hline\noalign{\smallskip}
QPS&  S & $\times$ & Rigetti & Java Script &S &\checkmark &\checkmark & Web-based IDE developed over open source simulator, Gui based circuit designer\\ 
\hline\noalign{\smallskip}
%S.No. & Simulator &  Library- Toolkit & Open-Source & Developer & Programming Language& Key Features & Pre-build Algorithm & QML support & Year\\ \hline
IQS & S & \checkmark & Intel & C++, Python & S &\checkmark & $\times$ &Optimally designed Quantum Circuit simulator for multi-core and multi-node architecture, uses message passing interface, Only qubits are available for exploration, other are abstracted \\ 
\hline\noalign{\smallskip}
Cirq &  B & \checkmark & Google & Python &B &\checkmark &\checkmark &Open source python based simulation library, supports both pure and mixed state simulation, Noise models supported, Also support external simulators \\ 
\hline\noalign{\smallskip}
Qiskit & B & \checkmark & IBM & Python & B &\checkmark &\checkmark&Set of simulators, Simulate gate and wave model, Noise model available, High-performance Cloud-based simulation \\ 
\hline\noalign{\smallskip}
QDK & B & \checkmark & Microsoft  & Q\# &B&\checkmark& $\times$ &QDK provides a development environment and provide support for a variety of Simulator as Full and Sparse state, Trace based resource estimators, Toffoli and noise simulators, Full state simulator supports up to 30 qubits  \\ 
\hline\noalign{\smallskip}
CuQuantum & B & \checkmark & NVIDIA & C, Python & S & $\times$ & $\times$ &Accelerated Computing for quantum simulation. C API as well as a wrapper for Python API, Supports the tensor network as well as state vector simulation,  It can integrate with different cloud computing platforms such as Cirq, Qiskit, and Pennylane.   \\ 
\hline\noalign{\smallskip}
Penny Lane& B & \checkmark & Xanadu & Python  & C & \checkmark &\checkmark&Cross Platform Python Library, Supports new programming paradigm for QC known as differential programming, Specially designed for QML, Available on amazon braket, Xanadu cloud platform\\ 
\noalign{\smallskip}\hline
Amazon braket & B &$\times$&Amazon &Python&C &\checkmark&\checkmark&Amazon braket provides 3 set of cloud-based simulator names as DM-1(Density Matrix),SV-1 (State vector),TN-1(Tensor Network), Performance is scalable as per the requirement\\
\noalign{\smallskip}\hline
% Quibo & S & \checkmark&-&  & S & \checkmark &\checkmark & Full Stack API for simulation, It also provides H/W support.\\
% \hline
\multicolumn{9}{c}{*S- Simulator, B- Both, Simulator \& Quantum Computer; ** S- Standalone application, C- Cloud Service B- Both}\\
\noalign{\smallskip}\hline
\end{tabular}
%% can be extended with LIQui, Quark, Scaffold 
\end{table*}
% \end{landscape}

 \item simulator\_stabilizer: It is a simulator based on Clifford circuits. Noise can be simulated if they are represented as Clifford gates \cite{bravyi2016improved}. It has a capacity of 5000 qubits. 
    \item simulator\_extended stabilizer: It is a simulator based on ranked-stabilizer decomposition, \textit{i.e.,} into a sequence of one or two-qubit Clifford gate. This decomposition is then used to predict the output of a quantum circuit \cite{bravyi2019simulation}. Non-Clifford gates in the circuit decide the stabilizer terms.
\end{itemize}
\subsubsection{IBM QCC}
\begin{figure}[t]
  \centering
  \includegraphics[scale=0.56]{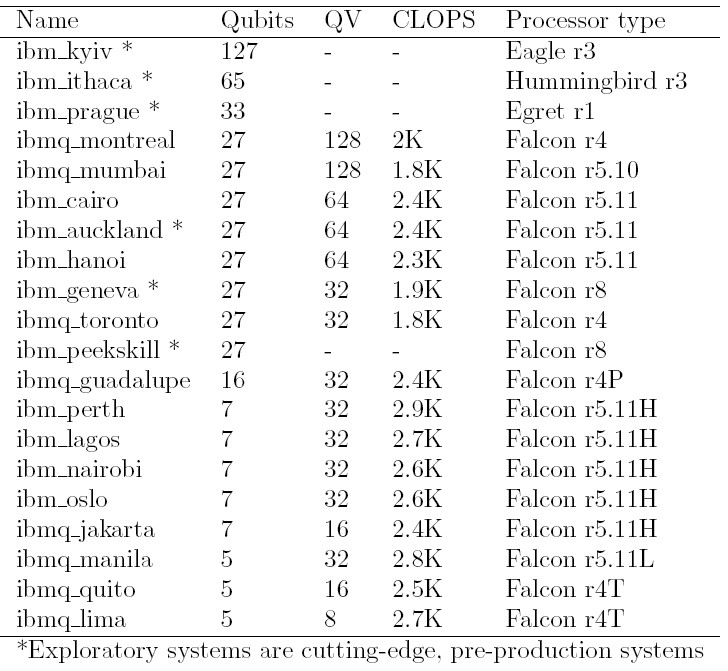}
  \caption{IBM Quantum Computers \cite{ibm_compute}}
  \label{fig:IBMQPU}
\end{figure}

IBM is one of the leading QCC service providers via IBM quantum lab. It provides a variety of computing resources for SC-QC. These systems are denoted by names that start with ibmq\_\* (older systems) or ibm\_\* (newer systems) associated with the city in which it is deployed. The details of the quantum compute resources available in the IBM quantum lab are shown in Fig. \ref{fig:IBMQPU}. The list of available hardware contains processors with 5 to 127 qubits, out of which up to 7 qubits are free and available for academia and industry research purposes. 

\begin{figure}[b]
  \centering
  \includegraphics[scale=0.45]{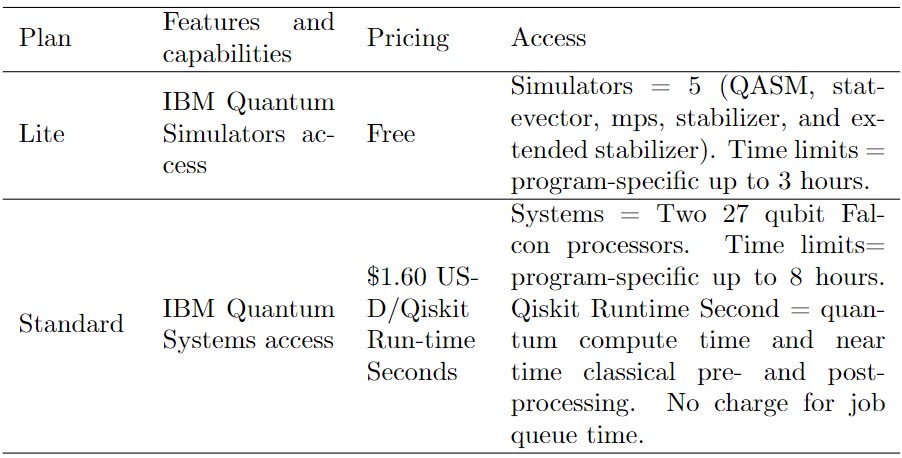}
  \caption{The Qiskit pricing models}
  \label{fig:Qpricing}
\end{figure}
\paragraph{IBM-QC} IBM has developed different processors for including Eagle, Hummingbird, Falcon and Canary as described below:
\begin{itemize}
    \item Eagle: It is a 127-qubit processor. Currently, it is available as two systems ibm\_washington and ibm\_kyiv. It incorporates more scalable packaging technologies so that the signals pass through multiple chip layers for high-density I/O without sacrificing performance. It supports up to 850 CLOPS. 
    \item Hummingbird: The hummingbird family of processors is 65 qubits QC, which utilizes a hexagon qubit layout which provides very few connections between qubits. 
    \item Falcon: Falcon processor family is useful for medium-scale circuits. It is a 27 qubits machine with 128 quantum volume and up to 2K flops. Falcon processors uses square topology and provides high quantum volume. 
    \item Canary: This processor family is useful for small-scale circuits. It contains processors from 5-16 qubits machines. It utilizes the 2D lattice arrangement for the qubits.  
\end{itemize}

\paragraph{QCC architecture} To access QCC services, a provider is used, which is a collection of hubs, groups, or projects. The hub may belong to an organization, and groups belong to an organization where a group can be associated with one or more projects. A public account by default relates to \textit{IBM-q/open/main} service provider. 
 % To get the details of the hardware and provider, the qiskit commands are utilized, as shown in the figure. 
% \begin{tcolorbox}
% \begin{tiny}
% \begin{verbatim}
% from qiskit import IBMQ
% IBMQ.load_account()
% provider = IBMQ.get_provider(group='open',
%                              project='main')
% system = provider.get_backend('ibmq_vigo')
% system.configuration()
% \end{verbatim}
% \end{tiny} 
% \end{tcolorbox}

\begin{figure}[ht]
  \centering
  \includegraphics[scale=0.32]{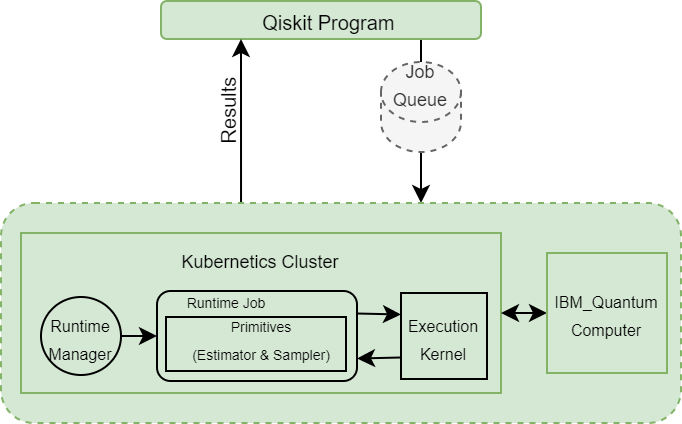}
  \caption{Qiskit run time architecture \cite{Qiskitruntime}}
  \label{fig:Qrt}
\end{figure}

Qiskit run-time as shown in Fig. \ref{fig:Qrt}, is a QCC service that optimizes the user workload for execution on a quantum computer. It includes additional primitives as a sampler and  estimator for better management of user workload. A sampler converts a user circuit to an error-mitigated circuit whereas an estimator gives the user to create a grouping of circuits and observable to better estimation of a parameter and its effect. IBM uses the fair share policy for providing services to all the jobs submitted and queued.

\paragraph{Qiskit Visualization} Qiskit provides a variety of visualization tools to analyze the results as well as understand the qubits and operations on them. Histogram plot, state plot, Bloch sphere, and Q-sphere are tools available for Qiskit visualization. The histogram plot is the standard plot for result analysis since it plots with the corresponding number of executions. The Bloch sphere is the standard single qubit sphere representation whereas Q-sphere is the improved version of the same. The difference between a Bloch sphere and Q-sphere is that the Bloch sphere gives a local view of the qubit state whereas the Q-sphere gives the global view of the qubit register on applying a quantum circuit. The Q-sphere simulation is only possible for five (or lesser) qubits. 

Limitations of Qiskit: Qiskit is one of the most feature-rich platforms but it's worth mentioning the limitations which helps to identify the gaps with other software. 
\begin{itemize}
    \item Quantum hardware access: It is one of the leading QCC providers with free access to five qubit machines and above is a pay-as-you-go basis. But other platform support is limited so different QC can not be accessed directly from Qiskit. 
    % It also provides access to PennyLane QC but not others.
    \item Run time limitations: The quantum provides two types of plans open and premium. The system limit with job execution is one hour for open plans and three hours for premium plans.  
    \item Learning Curve : Qiskit provides a very good community support as well as good tutorials and documentations still the learning curve is high for qiskit due to inherent complexity of QC. 
    
\end{itemize}
 
\subsection{Pennylane}
\begin{figure}[ht]
  \centering
  \includegraphics[scale=.32]{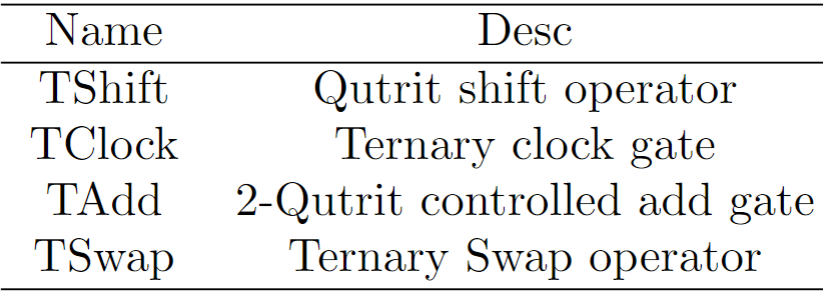}
  \caption{Qutrit Gates}
  \label{fig:Quitrit}
\end{figure}

PennyLane \cite{penny}  is an open-source software library for programming QC, developed and managed by Xanadu Quantum Technology. It provides a unified interface for training and deploying QML models on different quantum backends, including simulators and quantum processors. It is built on top of existing QC frameworks such as Qiskit and uses a similar interface to that of classical ML libraries like PyTorch and TensorFlow. It adapts the principle of differential programming in which the algorithms learn some parameters such as weights to train a QML model. Quantum circuits can also be programmed using a differential approach since quantum circuits are capable of self-adjusting the parameters (such as VQE circuits), which can be trained to learn as ML \cite{bergholm2018pennylane}. 
%copied%
\\
Such circuits can also be used in different domains such as quantum simulation or optimization. It can be fruitful in the development of quantum algorithms, the discovery of QEC, and the realization of physical systems. Along with the differential programming approach, it provides all the basic QC primitives, and it also provides seamless integration with the existing ML libraries such as TensorFlow and PyTorch.
\paragraph{PennyLane Features} It is a feature-rich platform with the following key features as described follows: 
\begin{itemize}
    \item Automatic differentiation: It supports inbuilt automatic differentiation of quantum gates which helps in parameter tuning for machine learning and optimization.
    \item Hybrid model support: It helps where classical ML libraries such as PyTorch, NumPy, and TensorFlow can be connected to quantum processors. 
    \item Optimization: Along with QML and QDL, it provides support for optimization problem solutions. 
    \item Cross-platform: It supports other platforms such as Cirq, Braket, and Qiskit with a help of a plugin, and the model/quantum circuit can be executed on different QC. 
\end{itemize}
This key feature describes the significance of PennyLane in QML. It also contains all the basic primitives of QC as well as advanced components. It supports all quantum operators such as gates, noisy channels, state preparations, and measurements. All of them have been discussed in detail.
\begin{itemize}
    \item Gates: It provides support for all the standard gates along with various parameterized gates such as rot, rotx or controlled rotx, etc. It supports various specialized gates such as quantum chemistry gates or gates constructed via the matrix, etc. 
    \item Noise modelling: It provides support for different noise channel modelling such as AmplitudeDamping, PhaseDamping, Depolarizing Channel, BitFlip, PhaseFlip, PauliError, ThermalRelaxationError, etc. 
    \item Qutrit operator: A qutrit operator which is a 3-level quantum system which is also supported as shown in Fig. \ref{fig:Quitrit}. 
\end{itemize}

% \begin{table}[ht]
% \caption{Qutrit Gates}
% \label{tab:tab3}
% \centering
% \begin{tabular}{m{3cm}m{4cm}}
% \hline
% Name & Desc  \\
% \hline

% TShift & Qutrit shift operator \\
% TClock & Ternary clock gate\\
% TAdd   &  2-Qutrit controlled add gate\\
% TSwap  & Ternary Swap operator\\

% \hline
% \end{tabular}
% \end{table}

\begin{figure}[ht]
  \centering
  \includegraphics[scale=0.55]{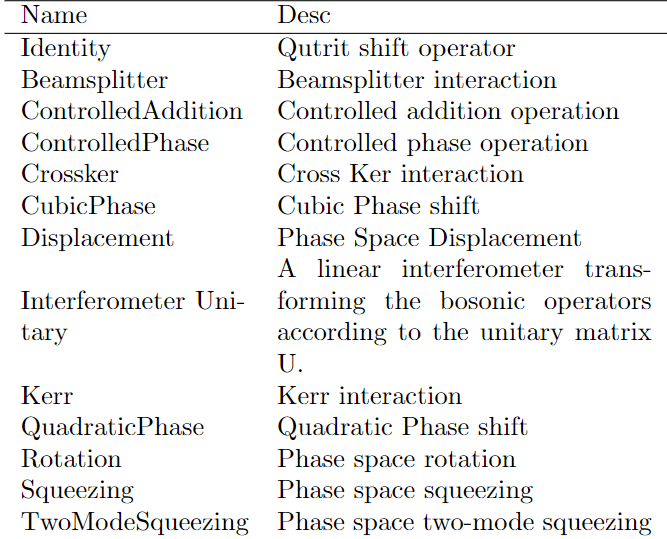}
  \caption{Continuous variable gates}
  \label{fig:CV}
\end{figure}

\paragraph{Continuous variable models} Continuous-variable quantum photonic circuits follow the continuous physical systems which reside in an infinite dimensional Hilbert space. Such quantum circuits follow continuous spectra operating on qumode. In contrast with the qubit, a qumode $\ket{\Phi}$ is given by the Eq. \eqref{eq:qumode} : 
\begin{equation}
   \label{eq:qumode}
   \ket{\Phi}  = \int dx~\Phi(x) \ket{x}.
\end{equation}
PennyLane supports the continuous variable model gates such as identity, beamsplitter, a cubic phase, kerr, etc. The details are shown in Fig. \ref{fig:CV}.
% \begin{table}[ht]
% \caption{Continuous Variable Gates}
% \label{tab:tab4}
% \centering
% \begin{tabular}{m{3cm}m{5cm}}
% \hline
% Name & Desc  \\
% \hline

% Identity & Qutrit shift operator \\
% Beamsplitter & Beamsplitter interaction \\
% ControlledAddition & Controlled addition operation \\
% ControlledPhase & Controlled phase operation\\
% Crossker & Cross Ker interaction\\
% CubicPhase & Cubic Phase shift\\
% Displacement & Phase Space Displacement \\
% InterferometerUnitary & A linear interferometer transforming the bosonic operators according to the unitary matrix U. \\
% Kerr & Kerr interaction\\ 
% QuadraticPhase & Quadratic Phase shift\\ 
% Rotation& Phase space rotation \\
% Squeezing& Phase space squeezing\\
% TwoModeSqueezing & Phase space two-mode squeezing\\

% \hline
% \end{tabular}
% \end{table}

\subsubsection{Quantum Simulator}
PennyLane library is designed in such a way that it can provide support for different hardwares without any compatibility issues. It can also support computation running across multiple quantum devices from different vendors. Even being a hardware-friendly platform it also supports the following quantum simulators:
\begin{itemize}
    \item default.qubit: It is a simple python based SV-simulator. It is useful in various use cases such as optimization with a significant number of qubits. It includes different QML back end such as Torch and TensorFlow. 
   \item default.mixed: It is a mixed-state quantum simulator useful in noise simulation and quantum channels. 
   \item default.Gaussian: It is a photonic simulator that operates on continuous variable quantum logic gates. 
   \item lightning.qubit: It is an SV simulator for faster simulation as compared to the \textit{default.qubit}. It is written in C++.
   \item lightning.gpu: It provides a GPU-accelerated quantum simulator utilizing NVIDIA GPU devices. It is an SV simulator that uses the cu\_QUANTUM SDK from NVIDIA which makes it suitable for simulating large qubits systems. 
\end{itemize}
The Pennylane provides the access to the QC such as D wave as well as support for an external plugin for Qiskit, Strawberry fields, etc. to connect it with different quantum cloud services. 

\subsubsection{D wave QCC}
D wave is the leading tech company in quantum developments with their most advanced annealing-based QC. It provides enterprise-grade solutions for business and scientific use cases that follow the most natural phenomenon of ground-state energy. The Annealing-based system tries to solve optimization problems and the problems that can be mapped to optimization problems via QC. The most advanced annealing QC with over 5000 qubits and 15-way qubit connectivity Advantage\textsuperscript{TM} QC designed to solve most complex business use cases. Leap is a software package that can be used to access and program Advantage QC. Leap is the only real-time quantum service designed to solve business problems. 

% \paragraph{Quantum  Annealing Principles }
% D wave system utilizes the process of Quantum Annealing which provides the minimum value in the search space based on the natural phenomenon of any system to always attempt to reach the lowest energy state. So, quantum annealing is crafted for optimization and problem which can be represented as optimization. For solving the optimization problem the annealing process can simultaneously initiate a multidimensional search with the help of a superposition search. Along with this quantum, tunneling can help bypass the local minima traps and reduces the search time. Entanglement helps in identifying the most suitable path for global minima. 

\paragraph{Advantage quantum computer}
The Advantage QC is currently in the 5\textsuperscript{th} generation of development with 5000 qubits, 3500 couplers, and 15-way qubit connectivity. It is far more power full to its predecessor processor D-wave-2000Q which was having 2000 qubits. The quantum advantage processor is available via ocean SDK as well as the leap hybrid solution by D-wave.

% It follows the pegasus topology as shown in Fig. \ref{fig:pegasus }. 
% \begin{figure}[h]
%   \centering
%   \includegraphics[ width=\linewidth,height=4cm]{Pegasus_qubits.png}
%   \caption{ A cropped view of the Pegasus topology with qubits represented as horizontal and vertical loops. This graphic shows approximately three rows of 12 vertical qubits and three columns of 12 horizontal qubits for a total of 72 qubits, 36 vertical and 36 horizontal. }
%   \label{fig:pegasus }
% \end{figure}

Limitation of Pennylane: Pennylane is a library specially designed for running programs on QCs but it also supports quantum simulations. It is based on differentiable programming and is best suited for QML programming. Still some limitation are found with this tool described below. 
\begin{itemize}
     \item Learning Curve: It is specially focused on QML so it will take lesser time to develop a quantum solution in Penny Lane but learning PennyLane will not be able to provide a good understanding of all QC concepts and their implementation. 
     \item Gate-Based implementation: Pennylane is specially designed for continuous variable computation. The gate-based system execution is a plugin based so it is less efficient as compared to other tools with direct support for universal gate-based computation.  
     \item Limited Community support: Pennylane is relatively new so it is having limited community support.  
\end{itemize}

\subsection{Microsoft-Quantum Development Kit (QDK)} It is an open-source Q-SDK development over Azure quantum created by Microsoft. It supports the Q\# a high-level development language for QC that have similar features such as C\#. It includes a simulator as well as a resource estimator. QDK provides a ready-to-use library and samples for different applications such as quantum chemistry, ML etc. QDK is more useful for developers familiar with Microsoft development tools such as visual studio and C\#. QDK also supports cross-development using Qiskit or Cirq. It provides support for optimization problem that utilizes classical as well as accelerated computing resource. Its Quantum inspired optimization tool provides different solutions such as parallel tempering, simulated annealing, population annealing, quantum Monte Carlo, etc.\\ 

\begin{figure}[ht]
  \centering
  \includegraphics[scale=0.40]{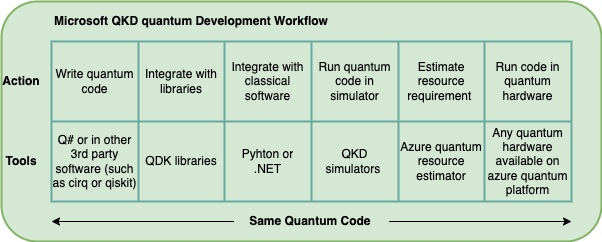}
  \caption{Microsoft quantum development kit-workflow  \cite{azure}}
  \label{fig:mqdk}
\end{figure}
\subsubsection{Quantum Simulator}
The Microsoft-QDK offers multiple quantum simulators, which allows developers to test and optimize their quantum algorithms using different methods. Details of the available simulators are provided as follows:   
    \paragraph{Full State Simulator} It is gate based SV simulator that can run on the local machine with a capacity of 30 qubits. It can be used to generate, test, and debug the universal gate model-based quantum circuits. 
    \paragraph{Sparse State Simulator} It is different from the SV simulator since it uses the sparse representation for the quantum state, which helps in the minimization of the memory requirements. A quantum system with sparse non-zero amplitudes on a computational basis can be simulated using the sparse state simulator. It can also help in simulating more capacity quantum systems. 
    \paragraph{Trace Based Resource Estimator} It is a resource estimator for executing a quantum program without actually creating the state. In this way, it can simulate more than 1000 qubits. It actually helps in estimating the resource required for a quantum program before actually executing it. 
    \paragraph{Toffoli Simulator} The Toffoli circuit can simulate only those circuits that uses Pauli-X and  CNOT  gates. Since Toffoli gates are assumed to be universal, quantum circuits can be converted to their Toffoli version and can be simulated using the Toffoli simulator. 
    \paragraph{Noise Simulator} The noise simulator helps to model the effects in the result when the noise occurs due to interaction with the environment.

%Envoking a simulator in QDK. 
\subsubsection{Azure Quantum}
Similar to amazon braket, Azure quantum also provides all the compelling QC devices as services. Following is the list of the QCC available on the Azure quantum platform :
 \paragraph{Quantinuum H1} Quantinuum H1 is a trapped ion QC designed by Honeywell. It follows the fully connected topology. It also supports mid-circuit measurement. 
    \paragraph{QCI}: A superconducting QC designed by quantum circuit inc. They are fast and high-fidelity QC with scalable architecture and modular design.
       
    \paragraph{IQloud} It is a specially designed quantum service for optimization problems. It also supports a modular architecture, and the optimization solution is crafted to solve industry-specific problems. 
    \paragraph{Microsoft QIO} This optimization solver is designed by Microsoft utilizing quantum principles. 
    \paragraph{SQBM+} SQBM+ is a quantum-inspired optimization solution based on the Simulated Bifurcation machine that is a combinatorial optimization solver utilizing the Simulated Bifurcation algorithm developed by Toshiba Corporation.
    \paragraph{IonQ} The IonQ systems are the same as those provided by the AWS service.
    \paragraph{Pascal} Pascal is a neutral atoms-based supercomputer. 
    \paragraph{Rigetti} The Rigetti systems are the same as those provided by the AWS service.

Limitation of QDK: It provides a quantum development platform as well as a plethora of hardware support through azure quantum. The following are the limitation of the QDK: 
\begin{itemize}
    \item Learning Curve: The learning curve for QDK is very high due to involvement of new programming language and lack of tutorial and active community. A language like Python is more handy for the developer due to wide usage as well easy learning. 
    \item Community Support: It has limited community of developer and relatively new library due to the active developers are limited.  
    \item Development platform: The QDK uses a development platform that is suitable for programmers working on visual studio but not for others. Q\# is itself a new language which increases the learning complexity.
    \item Pricing:  The pricing of the Azure quantum is also expensive with 10\$ per hour but it also provides up to 1 hour of free simulation. 
\end{itemize}

\subsection{Amazon Braket}
Amazon Braket \cite{braket} is the quantum development platform that provides QC hardware, Braket-Python SDK and a quantum simulator as a cloud service. In addition to offering the Braket-Python SDK, the platform also supports quantum development using other frameworks such as Qiskit and PennyLane. It is a part of Amazon Web Services (AWS) and offers a variety of Q-SDK with the convenience and cost-efficiency of the AWS pricing model.
\subsubsection{BRAKET Simulation} It has three different types of simulators each with a dynamic resource allocation so their performance is scalable. It provides the following simulator as described below: 
\begin{itemize}
    \item SV-1: It is a universal SV-simulator with a capacity of up to 32 qubits. It supports the parallel execution of circuits with a maximum runtime of six hours. 
    \item DM-1: It is a universal DM-simulator with a capacity of up to 17 qubits. It also supports the parallel execution of circuits with a maximum runtime of six hours.   
    \item TN-1: It is a TN-simulator with a maximum capacity of up to 50 qubits. The TN-1 is useful in simulating sparse circuits and circuits with spatial features such as QFT. 
\end{itemize}
The simulators provided by Braket are useful in various scenarios, as outlined in the table \ref{tab:amzsim}.
\begin{table}[ht]
\caption{Braket Quantum Simulator Use Case Comparison}
\label{tab:amzsim}
\centering
\begin{tabular}{m{2cm}m{1cm}m{1cm}m{1cm}m{1cm}}
\hline
\textbf{Use Case} & \textbf{Local} & \textbf{SV1} & \textbf{TN1 } & \textbf{DM1}  \\
\hline
Debugging and Prototyping& \checkmark & - &- &-\\
Large Scale Experiment&- & \checkmark &\checkmark &\checkmark\\
Parallel Circuit Simulation&- &\checkmark &\checkmark &\checkmark\\
Noise Simulation&\checkmark & - &- &\checkmark\\
Special Structure Circuit simulation &- & - &\checkmark &-\\
\hline
\end{tabular}
\end{table}

% \begin{figure}[h]
%   \centering
%   \includegraphics[scale=0.22]{chart-of-simulator-use-cases.png}
%   \caption{Amazon Braket Quantum simulator comparison \cite{braketsim}}
%   \label{fig:amazonsim}
% \end{figure}

% \begin{table}[!ht]
%     \caption{Braket Pricing : Braket pricing option for different Quantum hardware per task and per shot basis} 
%     \centering
%     \begin{tabular}{llll}
%     \hline
%         Hardware Provider  & QPU family  & Per-task price  & Per-shot price  \\ \hline
%         IonQ  & IonQ device  & \$0.30000  & \$0.01000  \\ 
%         OQC  & Lucy  & \$0.30000  & \$0.00035  \\ 
%         Quera & Aquila & \$0.30000 & \$0.01000 \\ 
%         Rigetti  & Aspen-11  & \$0.30000  & \$0.00035  \\ 
%         Rigetti & Aspen-M & \$0.30000  & \$0.00035  \\ 
%         Xanadu  & Borealis  & \$0.30000  & \$0.0002  \\ \hline
%     \end{tabular}
% \end{table}

\begin{figure}[t]
  \centering
  \includegraphics[ width=\linewidth]{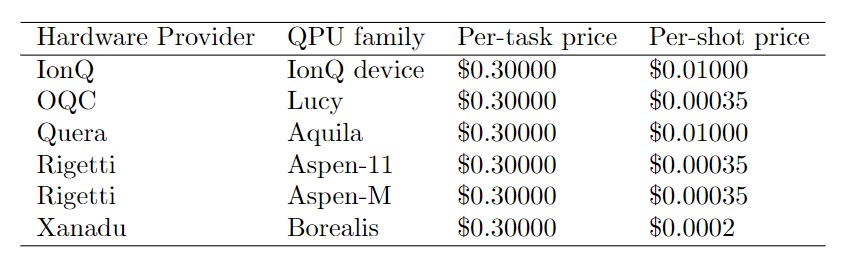}
  \caption{ Braket Pricing Braket pricing option for different Quantum hardware per task and per shot basis. Source: \url{https://aws.amazon.com/braket/pricing/} }
  \label{fig:pricing }
\end{figure}

\subsubsection{Amazon Braket QCC \cite{AWSQC} }
Amazon does not produce any QC hardware itself, however, it offers access to a range of quantum processors from various providers. The list of QC providers is as follows:  

    \paragraph{D-Wave}  D-wave's \cite{dwave} quantum annealing systems are available through amazon braket with real-time access to the systems such as D-wave-Q-2000 and D-wave-Advantage. It can be accessed via the ocean braket plugin or directly with braket SDK. The ocean plugin has the added advantage of a sampler choice of solution methods for the user. 
    % A problem is described using a set of values that represents the weights of qubits and the strength of couplers.%%copied 
    \paragraph{Ion-Q} Braket also provides IoN-Q \cite{ionq}, a TI-based QC utilizing ionized ytterbium atoms. IoN-Q systems are useful in chemistry, material sciences, and optimization. Trapped IoN-Q utilizes the fully connected topology. 
    % For programming Ion-Q processors, laser-based gate operation is applied. 
    \paragraph{OQC} Oxford Quantum Circuits (OQC) \cite{oxford} are SC-qubits based QC available on Braket. The OQC designed by oxford  has a proprietary ‘Coaxmon’ technology for its QC. The Coaxmon uses a 3D architecture for a control plane perpendicular to the qubit which is better in control circuitry as compared to the 2D lattice. The lucy-OQC system is an eight qubit-QC based on Coaxmon technology.  
    \paragraph{QuEra} QuEra a neutral atom-based QC available on Braket. It is a Rydberg atom qubit that utilizes rubidium ion. The operation/gates are applied using laser beams. 
    \paragraph{Rigetti} Rigetti QC \cite{rigetti} are gate-based devices using the SC-qubits. Its Aspen series processors have tileable lattices that support scalable architecture.
    \paragraph{Xanadu}  Xanadu’s QC \cite{xanadu} are photonic-QC that implements photons-based qubits via laser operations. It also supports continuous-variable QC. They are also available via PennyLane quantum service. Borealis is the first photonic QCC available through Amazon Braket.    
% \begin{figure}[ht]
%   \centering
%   \includegraphics[ width=\linewidth]{Images/20_borealis.png}
%   \caption{Borealis-216 squeezed-state qubits photonic quantum processor from Xanadu \cite{Borealis}}
%   \label{fig:borealis}
% \end{figure}

Limitations of Amazon Braket: Amazon bracket is a very rich and diverse platform in terms of support of different QC and simulators. Still, the following are the considerable limitations of the amazon braket platform. 
\begin{itemize}
    \item Compute resource access: The amazon braket platform provides a choice of all leading QC as a service but it does not provide any Quantum platform developed by amazon itself. The simulator choices are limited in braket.
    \item Pricing:  The AWS braket provides one hour of free simulation per month for free tier users which is not sufficient for learning. The pricing is very precise so for a new learner it might become very expensive due to vague test cases and designs.
    \item Limited Control: It provides limited control over the hardware since all the hardware are accessed via an API. This may limit the ability to optimize circuits and experiment with different configurations.
\end{itemize}

Cirq, Qiskit, Pennylane, Braket, and QSDK are the most feature-rich quantum development platform. A detailed discussion has been provided with advantages and limitation is provided to create a comparison among these tools. Other than these tools there are other worth mentioning tools described further. The majority of them are stand-alone simulators developed in the initial stage of quantum development. But, these tools formed the basis for quantum development progress. 

%% useful url

% https://ionq.com/posts/june-24-2021-hello-many-worlds
\subsection{Other Quantum Tools} The list of the quantum simulator is long but not all of them are equally important. Some of the initial simulators are now outdated or not under an active development cycle. Due to the limited availability of resources and documentation, only a brief discussion can be provided for such tools. So, in the rest of subsection some notable quantum development tools are discussed that includes standalone simulator or under development projects.

    \subsubsection{cuQuantum}  cuQuantum \cite{cuq}  is a Q-SDK  developed by NVIDIA including quantum simulator based on python and Q\#. It consists of optimized libraries and a toolkit for quick implementation of a quantum algorithm. It also provides a way to run quantum algorithms on NVIDIA GPUs and Tensor processor that can accelerate the simulation of large-scale quantum circuits. It is built on the top of Cuda platform so it can be easily used with any NVIDIA GPU. Currently it is only available for linux platform. 
    % NVIDIA is attempting to provide accelerated computing to quantum simulations. 
    % Using Tensor Processing Unit(TPU) and Graphics Processing Unit(GPU), quantum simulations can be speed-up. 
    It supports two types of simulations, \textit{i.e.}, SV and TN-simulation as follows: 
    \begin{itemize}
        \item cuStateVec: A specially designed high performance library for SV-simulation with API supporting quantum gates, measurement, sampling and qubit reordering, multi GPU computation \cite{custate} .  
        \item cuTensorNet: A high performance library for the TN computation by providing efficient calculation of tensor contractions\cite{cutensor}.
    \end{itemize}
It also provides Appliance that is a container-based pre-installed library (cuStatevec, cuTensorNet, cuQuantum, and cuQuantum python API) with multi-GPU back-end support.

   \subsubsection{Forest } Forest is the Q-SDK developed by Regetti that includes Pyquil, quilc and qvm. Pyquils is a python library to provide support for quantum instruction language Quil. The main purpose is Quil based simulation and executing a Quil program over QC \cite{smith2016practical}.  

    \subsubsection{IQS} Intel-QS \cite{iqs} is an optimized universal gate-based quantum simulator optimized for the multi-core and multi-node architecture \cite{intelQS}. It was formally known as qHiPSTER. It utilizes a message-passing protocol for distributed resources. 

     \subsubsection{Leap Quantum cloud service} Leap is a service provided by D-wave for free and real-time access to the Advantage system. It is a feature-rich platform with access to both QC and CC resources, Ocean SDK, Hybrid solver, and demos. It also provides a constraint quadratic solver with the support of continuous variable QC. 
   
    \subsubsection{QuEST} QuEST(Quantum Exact Simulation Toolkit) \cite{jones2019quest} is a QC simulation toolkit capable of efficiently performing quantum circuits, SV and DM simulations. It is a high-performance simulator and can run on a variety of devices. It supports GPU acceleration and can also run in a distributed environment. 

    \subsubsection{Qx Quantum Computer Simulator} It is a simple quantum simulator that allows the user to simulate universal gate-based QC \cite{QX}. It also includes a quantum assembly language known as Quantum Code (QCode), which is a text-based representation of user-defined circuits and serves as input for the simulator. It supports quantum register, quantum to the classical interface, circuit splitting, and scheduling.

    \subsubsection{Quantum++} It is a quantum simulator written in C++ language \cite{Q++Gheorghiu2018}\cite{qpp} and requires no additional support. It can easily simulate 24 pure qubits (12 mixed states) qubits on an 8 GB machine. It supports all the standard quantum logic gates as well as classical reversible gates.
    
    \subsubsection{QPS}  Quantum Programming Studio (QPS) \cite{QPS} is a web-based quantum simulator and IDE. It is built on top of an open-source quantum simulator. It provides QPS clients with the help of that you can connect to local or cloud systems such as Rigetti QCS or IBM Qiskit.   

    \subsubsection{Strawberry Fields} It is cross development library for quantum simulation and executing algorithms on Xanadu's photonic processor. It supports other libraries as Penny Lane, TensorFlow for accelerated development. 

% After reviewing all the popular quantum development, It the time to move ahead in the quantum journey. The quantum development cycle is the way to understand how to develop a solution by utilizing the quantum development tools. So in the next section first the development cycle of a quantum solutions are discussed followed by implementation of building blocks. After studying the building blocks an end to end solution development is discussed with a comparision of 3 different tools. 
 
% \section{Plan in action: State of the art quantum research}
% \label{sec:5}
After exploring the details of the quantum development tools, the execution environment, and building blocks, We can now start the quantum journey by creating a pathway using the building blocks learned. So, the next section will help in utilizing these tools to complete the quantum journey. 
\section{The Quantum Journey: A Quantum Development life cycle }
\label{sec:5}
This section first outlines the four phases of QDLC, detailing the progression of the quantum development cycle in different phases. Then the building blocks along with the required tools are discussed. A toy example of ``Hello Quantum World" is also provided for illustration. Lastly, a full end-to-end tutorial of QML is presented using Qiskit and Penny Lane.
\subsection{Quantum Development Life Cycle}
\begin{figure}[t]
  \centering
  \includegraphics[width=\linewidth]{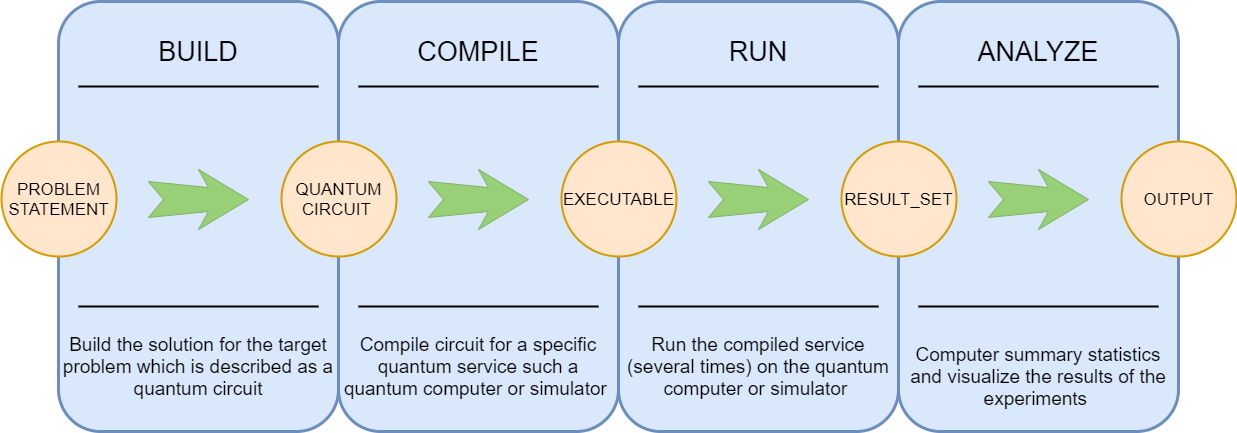}
  \caption{The quantum development life cycle}
  \label{fig:QDLC}
\end{figure}

 The life-cycle for quantum development presented in \cite{dey2020qdlc}\cite{weder2022quantum}, are inspired from the standard software development life cycle. These are not much helpful for NISQ-era devices, therefore we will be using a different QDLC that is simplified and more suitable for quick quantum development. It is shown in Fig. \ref{fig:QDLC} and abstracted from Qiskit\footnote{\url{https://qiskit.org/documentation/intro_tutorial1.html}} Textbook and Microsoft QSDK\footnote{\url{https://learn.microsoft.com/en-us/azure/quantum/overview-what-is-qsharp-and-qdk}}.
NISQ-era quantum solutions are being developed at the hardware level, with algorithms being implemented as quantum circuits. Usually, a QC requires a CC device for the data input and output. In general input data is encoded into qubits and superimposed to explore QC benefits and then the measurement is performed to get the output as bits. The QDLC will cover pure quantum algorithms and the quantum portion of hybrid algorithms. It consists of the following four phases, \textit{i.e.}, build, compile, run and analyze as described below: 
\begin{enumerate}[i).]
    \item Phase-I-Build: In this phase, first the QC-based solution for a given problem is developed as a quantum algorithm. Then the quantum algorithm is converted into a quantum circuit which is a collection of, simple gates or trainable gates (parameterized) as per the solution. The circuit can be designed by only using universal gates. Generally, QML solutions are designed to use trainable gates whose parameters are adjusted while iterative training. 
    The quantum circuit is then represented in a QC language (such as QML\footnote{It is a quantum programming language, not quantum machine learning.}, LANQ) before feeding into a simulator or a QC. A high-level language (such as python) can also be used to describe the circuit which is converted to low level using a compiler in Phase II. 
    \item Phase-II-Compilation:
     The output of phase I is used in phase II, \textit{i.e.,} a circuit description in a quantum programming language.
     This circuit program goes as an input to the quantum compiler \cite{maronese2022quantum}. The compilation process converts the circuit description to a low-level language or machine instruction executable suitable for a particular hardware. 
    
    \item Phase-III-Execution:
    The input data for the execution is provided with help of a CC. After providing the input data and machine-executable the system can execute the given circuit. The result of quantum circuits is probabilistic in nature as well as affected by noise, So  single execution can't be treated as a correct result. So, a circuit is executed several times (suppose 1024 times) and results are collected. 

    \item Phase-IV-Analysis: The result of several runs are collected and viewed as the probability distribution and high probability results are selected as the solution. In the analysis phase, the collected result is analyzed statistically or graphically to decide on the output. It also helps in making the decision if the experiment needs to be carried out again. 
\end{enumerate}

After looking at the QDLC, let's start with the first phase build and see how the build process is done from a given algorithm to a quantum circuit. First, let's look at the building blocks and their implementation. 

\subsection{Implementation of the building block}
The building blocks of quantum development are qubits, logic gates, and quantum circuit. Their implementation is provided for Cirq, Qiskit, and PennyLane. 

\subsubsection{Initialization of Qubits} The input data is provided in the form of qubits and all the operations take place on the qubit. The process of defining a qubit is described below:

\paragraph{Cirq} In Cirq, qubits are created and initialized simultaneously, by specifying their location on a one-dimensional or two-dimensional lattice respectively.\\
\begin{tcolorbox}[width=8cm]
\begin{tiny}
\begin{verbatim}
     qubit = cirq.GridQubit.square(2)
\end{verbatim}
\end{tiny}
\end{tcolorbox}
     The above syntax will create four qubits arranged in a square grid. Other options for qubit creation are linequbit and namedqubit, which are described below: 
     \begin{itemize}
    \item cirq.NamedQubit: It is used to provide abstract identification to qubits.
    \item cirq.LineQubit: It is used to create linear qubits array.
    \item cirq.GridQubit: It is used to create a rectangular grid of qubits.
\end{itemize}
    \paragraph{Qiskit} In Qiskit the qubits are not created separately but a circuit is initialized with the desired number of qubits. As in the example below: \\
 \begin{tcolorbox}[width=8cm]
\begin{tiny}
\begin{verbatim}
    qc = QuantumCircuit(1)\\
    initial\_state = [0,1]\\ 
\end{verbatim}
\end{tiny}
\end{tcolorbox}
    In the above code, a quantum circuit qc is created with one qubit. The initial state will initialize qubit 0 to 1. 
    
    \paragraph{Pennylane} In PennyLane qubits are not separately defined but it is in the form of qnode running over a device. 
  \begin{tcolorbox}[width=8cm]
\begin{tiny}
\begin{verbatim}
import pennylane as pl
dev1 = pl.device("default.qubit", wires=1)
\end{verbatim}
\end{tiny}
\end{tcolorbox}
In the above code, a device dev1 is created and it is assigned a qnode. A circuit is defined where the parameter will be a qubit. 

% @pl.qnode(dev1)
% def circuit(param):
%     pl.RX(param[0], wires=0) 
%     return pennylane.expval(pl.PauliZ(0))

% print(circuit([0.55]))

\subsubsection{Implementation of the Quantum logic Gates}
Quantum gates are supported by all Q-SDK with some variations. Table \ref{tab:gate} provides the implementation of different gates in Cirq, Qiskit \footnote{The Qiskit library is imported as qc using \textit{import Qiskit as qc}}
and Pennylane \footnote{Pennylane library is imported as pl using \textit{import Qiskit as pl}}. 

\begin{table}[ht]
\caption{Implementation of Quantum Logic Gates}
 \label{tab:gate}
\centering
\begin{tabular}{m{0.6cm}m{2.4cm}m{2cm}m{2cm}}
\hline
Gate & Cirq & Qiskit  & Pennlylane    \\
\hline
% \multirow{}{}{}
X & cirq.X(q) & qc.x(q) &pl.PauliX(q)\\
Y  &cirq.Y(q)  & qc.y(q)&pl.PauliY(q) \\
Z  &cirq.Z(q)  & qc.z(q)&pl.PauliZ(q)\\
H &cirq.H(q) &qc.h(q)&pl.Hadamard(q)\\
CNOT & cirq.CNOT($q_{0},q_{1}$) & qc.cx($q_{0}, q_{1}$)& pl.CNOT($q_{0},q_{1}$)\\ \hline
\end{tabular}
\end{table}

\subsubsection{Implementation of the Quantum Circuit}
A quantum circuit is a collection of logic gates connected via wires. It takes at least one qubit as input and one measurement gate before the output. Let's see how to implement a quantum circuit in different development platforms: 
\paragraph{Cirq} The following code is displaying a sample circuit implementation example in Cirq. 
 
\begin{tcolorbox}[width=8cm]
\begin{tiny}
\begin{verbatim}
q = cirq.GridQubit.square(3)
# The X gate is created.
PauliX_gate = cirq.X
# It is applied on qubit zero.
PaulitX_op = x_gate(q[0])
\end{verbatim}
\end{tiny}
\end{tcolorbox}
In the above code first, a square grid of nine qubits is created. Then a single qubit X gate is applied to the q[0].    

    \paragraph{Qiskit} In Qiskit the qubits can't be defined separately but a quantum circuit is directly defined with a number of qubits parameters. 
    %%copied

 \begin{tcolorbox}[width=8cm]
\begin{tiny}
\begin{verbatim}
from qiskit import QuantumCircuit
# Create a circuit with a register of three qubits
qc = QuantumCircuit(3)
# The Hadamard gate is applied to create 
#superposition of zero and one basis state.
qc.h(0)
#CNOT gate is applied where qubit 0 is used control qubit
#qubit 1 as the target bit.
qc.cx(0, 1)
\end{verbatim}
\end{tiny}
\end{tcolorbox}   

\paragraph{Pennylane}  In PennyLane, a quantum circuit is defined as a python function. 
  \begin{tcolorbox}[width=8cm]
\begin{tiny}
\begin{verbatim}

def circuit(param):
    pennylane.RX(param[0], wires=0)
    pennylane.RY(param[1], wires=0)
    return pennylane.expval(pl.PauliZ(0))
\end{verbatim}
\end{tiny}
\end{tcolorbox}

\subsubsection{Acquiring the Run time}: After the circuit is ready it needs to execute on a hardware or simulator. 
   \paragraph{Cirq} Cirq doesn't provide public access to the QC even though the method to connect to the QC is given below. The below code describes how we can get access to the QC using Cirq. It can be used to access other QCC (e.g. AQT pine-systems or ION-Q hardware) environments via their API.
\begin{tcolorbox}[width=8cm]
% title= Obtaining Quantum Computer: Cirq 
% \textbf{ }
\begin{tiny}
\begin{verbatim}
engine = cg.get_engine()
processor = engine.get_processor(processor_id)
device = processor.get_device()
print(device)
valid_qubit = sorted(device.metadata.qubit_set)[0]
# Transform circuit to use 
#an available hardware qubit.
hw_circuit = 
circuit.transform_qubits(lambda q: valid_qubit)
print(hw_circuit)

\end{verbatim}
\end{tiny}
\end{tcolorbox}  

The above code can be used for a public quantum computer by obtaining a device with the processor\_id provided. Any code needs to validate first and transform as per the qubit arrangement on the hardware. So, a transform method will convert the qubit arrangement as per the hardware qubit. 

% # Upload the program and submit jobs to run in one call.
% job = processor.run_sweep(
%     program=hw_circuit,
%     repetitions=1000)

% print("Scheduled. View the job 
% at: https://console.cloud.google.com/quantum/"
% "programs/{}?&project={}".format(job.id(), project_id))

% # Print out the results. This blocks until the results are returned.
% results = job.results()
% print("\nMeasurement results:")
% for result in results:
%     print(result)
Obtaining Simulator using Cirq: A simulator can be obtained from GCS using the following method.

\begin{tcolorbox}[width=8cm]
% [title = Obtaining Quantum Simulator: Cirq or QSim, width=8cm]
% \textbf{ }
\begin{tiny}
\begin{verbatim}
#importing the cirq and qsim library
import cirq
import qsimcirq 
#Executing using a simulator to obtain result
simulator = cirq.Simulator()
result = sim.run(circ, repetitions=5)

# Simulate the circuit with qsim 
qsim_sim = qsimcirq.QSimSimulator()
qsim_res = qsim_simulator.run(circ, repetitions=5)
\end{verbatim}
\end{tiny}
\end{tcolorbox}  
The above describes obtaining a quantum simulator for both Cirq as well as QSim. For simulation, the instance of the simulator is provided with the circuit to be executed as well a number of repetitions to be made.

% # Obtaing 4 qubit in a grid arrangement
% qubits = cirq.GridQubit.square(2)

% # Define a circuit to run
% circ = cirq.Circuit()
% circ.append([cirq.CZ(q2, q3), cirq.H(q1)])
% # Qubit Measurement for results
% circ.append(cirq.measure(*qubits, key='all_qubits'))

\paragraph{Qiskit} Obtaining the simulator requires the following set of codes.
\begin{tcolorbox}[width=8cm]
% title=Obtaining Quantum Simulator: Qiskit
\begin{tiny}
\begin{verbatim}
#library import for QASM simulator
from qiskit.providers.aer import QasmSimulator
from qiskit.providers.basicaer import QasmSimulatorPy
#obtaining instance of QASM simulator
qasmsim = QasmSimulator()

compiled_circ = transpile(circ, qasmsim)
job = qasmsim.run(compiled_circ, shots=1024)
result = job.result()
\end{verbatim}
\end{tiny}
\end{tcolorbox}

In the above code, the instance of Qasm simulator is obtained using the QasmSimulator(). After obtaining simulator the transpile() function will create a compiled circuit using the circuit and simulator instance. For running the simulation the run() method will be called with compiled\_circ and the number of repetitions as input parameters. 

\paragraph{PennyLane} PennyLane is designed in such a way that by default it will access the QC. The simulator can be obtained specifically by the following methods. It also supports the Qiskit simulator using the Pennylane-Qiskit API. 
\begin{tcolorbox}[width=8cm]
% title=Obtaining Quantum Simulator: Pennylane
\begin{tiny}
\begin{verbatim}
#obtaining inbuilt simulator 
dev0 = pl.device('default.qubit', wires=1)

import cirq
#obtaining a cirq simulator
sim1 = cirq.Simulator()
dev1 = pl.device('sim1', wires=2)

#This device dev can be executed as any other pennylane devices
#obtaining Qiskit simulator
from qiskit.providers.aer import QasmSimulator
sim2 = QasmSimulator()
dev2 = pl.device('sim2', wires=2)
\end{verbatim}
\end{tiny}
\end{tcolorbox}
In the above code, the method is explained for obtaining three different simulators as d\textit{efault.qubit, cirq} and \textit{qiskit}. The simulator is obtained while creating a device instance. Inbuilt simulators are directly obtained while for external simulators a library is required to be imported. As in the above code \textit{sim1} is an instance of the Cirq simulator while \textit{sim2} is an instance of the Qiskit simulator. 

\subsection{ ``Hello Quantum World": A Toy Example of Quantum Programming}
In quantum programming a ``hello world" example is taken as a simple program for establishing a bell state between two qubits also known as entanglement. Quantum entanglement correlates two qubits in such a way that both behave as a system. For entangled pair of qubits, changes in one qubit will be instantly reflected in the other qubit even after separating them apart. On measurement both will attain either $\ket{0}$ or $\ket{1}$ state. There is a total of four bell states out of one state that will be used for the toy example. The bell state is represented as shown in Eq. \eqref{eq:bellstate}
\begin{equation}
\label{eq:bellstate}
|\Phi^+\rangle = \frac{1}{\sqrt{2}} (|0\rangle_A \otimes |0\rangle_B + |1\rangle_A \otimes |1\rangle_B) \\    
\end{equation}

\begin{equation}
|\Phi^-\rangle = \frac{1}{\sqrt{2}} (|0\rangle_A \otimes |0\rangle_B - |1\rangle_A \otimes |1\rangle_B) 
\end{equation}

\begin{equation}
|\Psi^+\rangle = \frac{1}{\sqrt{2}} (|0\rangle_A \otimes |1\rangle_B + |1\rangle_A \otimes |0\rangle_B) \\           
\end{equation}

\begin{equation}
|\Psi^-\rangle = \frac{1}{\sqrt{2}} (|0\rangle_A \otimes |1\rangle_B - |1\rangle_A \otimes |0\rangle_B) \\    
\end{equation}
\begin{figure}[ht]
  \centering
  \includegraphics{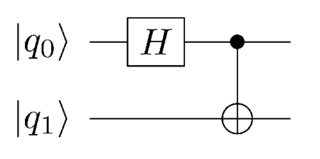}
  \caption{Circuit Diagram for Entanglement Creation}
  \label{fig:EPR}
\end{figure}

    \paragraph{Cirq} As the quantum algorithm is implemented in the form of a  quantum circuit, the corresponding circuit for bell creation is shown in Fig. \ref{fig:EPR}. The EPR creation code is provided below. 
    \begin{tcolorbox}[colback=black!5!white,colframe=white!5!black,title= Bell State: Cirq,width=8.5cm]
% \textbf{Cirq Example of Bell State Creation}
\begin{tiny}
\begin{verbatim}

1.  import cirq_google #library import
2.  import matplotlib.pyplot as plot
# Create a circuit instance as qc 
3.  qc = cirq.Circuit() 
4.  q_0, q_1 = cirq.LineQubit.range(2)
5.  qc.append(cirq.H(q_0))
6.  qc.append(cirq.CNOT(q_0, q_1))
# Obtaining and initializing the simulator

7.  s = cirq.Simulator()
8.  res = s.simulate(qc)
9.  print(res)

# Measurement gate is used to obtain the output
10. qc.append(cirq.measure(q_0, q_1, key='res'))

# Sample the circuit
11. results = s.run(qc, repetitions=1024)

12. cirq.plot_state_histogram(results, plot.subplot())
13. plot.show()
\end{verbatim}
Source: \url{https://www.quantumai.google/cirq/start/basics}
\end{tiny} 
\end{tcolorbox}

\begin{figure}[ht]
    \centering
    \includegraphics[scale=0.45]{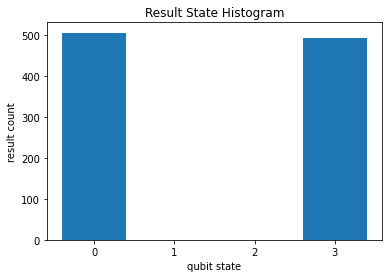}
    \caption{Result: Out of 1000 execution, 50\% of the execution achieved state 0 by both the qubits  q\textsc{0}       q\textsc{1} \textit{i.e.} ($\ket{00}$) and  50\% state 1 is achieved by both qubits q\textsc{0} q\textsc{1} \textit{i.e.}     ($\ket{11}$}
    \label{fig:outcirq}
    \end{figure}

The above Cirq implementation is for the first bell state $\Phi^+$. In the code first line is used to import the cirq library. lines 2-5 will implement the required circuit. Line 6 will acquire a simulator and line 11 will execute the same circuit 1024 times. The output is shown below and  lines 13-14 will help in visualizing the results shown in Fig. \ref{fig:outcirq}. In the output, it is clearly visible that the bell state of $\sqrt{(1/2)}(\ket{00}+\ket{11})$ is obtained. And the histogram shows 50-50 percent attainment of both state 0($\ket{00}$) and state 3($\ket{11}$). 

\begin{tcolorbox}[colback=black!5!white,colframe=white!5!black,title=Output,width=8cm]
\begin{tiny}
qubits: (cirq.LineQubit(0), cirq.LineQubit(1))
output vector: 0.707|00⟩ + 0.707|11⟩
phase:
output vector: |⟩
    
\end{tiny}

\end{tcolorbox}

\begin{figure}[ht]
  \centering
  \includegraphics[scale= 0.4]{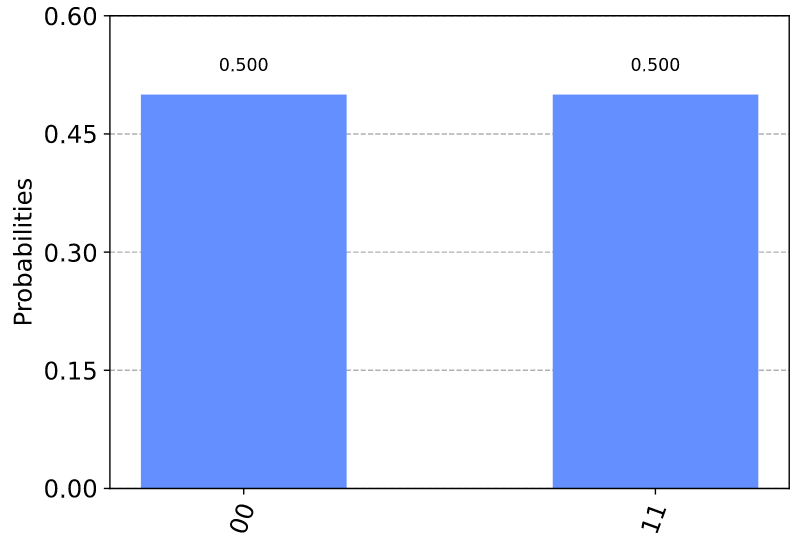}
  \caption{Result:Out of 100 execution, 50\% of the execution achieved state 0 by both the qubits  q\textsc{0} q\textsc{1} \textit{i.e.} ($\ket{00}$) and  50\% state 1 is achieved by both qubits q\textsc{0} q\textsc{1} \textit{i.e.} ($\ket{11}$}
  \label{fig:outQiskit}
\end{figure}

\paragraph{Qiskit} The Qiskit implementation of the Bell state and its output is shown below.
    \begin{tcolorbox}[colback=black!5!white,colframe=white!5!black,title= Bell State: Qiskit,width=8.5cm]
% \textbf{Qiskit Example of Bell State Creation}
    \begin{tiny}
    \begin{verbatim}
1.  from qiskit import QuantumCircuit, Aer, assemble
2.  from qiskit.visualization import plot_histogram, 
3.  from qiskit.visualization import array_to_latex
4.  from qiskit.visualization plot_bloch_multivector
5.  import numpy as np

6.  qc = QuantumCircuit(2)
7.  qc = QuantumCircuit(2) 
# Apply hadamard gate on the qubit 0
8.  qc.h(0) 
# Apply CNOT gate on the qubit 0,1
9.  qc.cx(0,1)
10. qc.draw()

# Let's get the result:
11. qc.save_statevector()
12. qobj = assemble(qc,shots=1000)
13. result = svsim.run(qobj).result()
# Print the state vector neatly:
14. final_state = result.get_statevector()
15. array_to_latex(final_state, prefix="\\text{Statevector = }")
16. plot_histogram(result.get_counts())

\end{verbatim}
Source: \url{https://www.qiskit.org/textbook/ch-gates/multiple-qubits-entangled-states.html}
\end{tiny}
\end{tcolorbox}
The above Qiskit implementation is similar to the Cirq-EPR example. In this, lines 1-5 are used to import the necessary libraries. The lines 6-10 will implement the required quantum circuit. Line 11 is used to save the state vector. Lines 12-13 is used for execution in which line 11 is initializing the simulator with input as quantum circuit and the number of shots is 1000. Line 15 is displaying the state vector and line 16 is used for plotting the histogram as shown in Fig. \ref{fig:outQiskit}. 

\begin{tcolorbox}[colback=black!5!white,colframe=white!5!black,title=Output,width=8cm]
\begin{tiny}
\text{Statevector = }
$\begin{bmatrix}
\sqrt{(1/2)}(\ket{00}+\ket{11}
 \end{bmatrix}$
\end{tiny}
\end{tcolorbox}

\paragraph{Pennylane} Implementation of bell state creation using penny lane is shown below. 

\begin{tcolorbox}[colback=black!5!white,colframe=white!5!black,title= Bell State: PennyLane,width=8.5cm]
% \textbf{Qiskit Example of Bell State Creation}
\begin{tiny}
\begin{verbatim}
#Program to implement the bell state in PennyLane

#Library Imports

1.  import pennylane as pl
2.  from pennylane import numpy as pl_np

#Acquiring  a device for execution
3.  dev1 = pl.device("default.qubit", wires=2, shots=1024)

4.  @pl.qnode(dev)
5.  def circuit():  # Defining the Circuit
6.      pl.Hadamard(wires=0)   
7.      pl.CNOT(wires=[0, 1])
8.      return pl.sample(pl.PauliZ(0)), pl.sample(pl.PauliZ(1))
\end{verbatim}
\end{tiny}
\end{tcolorbox}
In bell state example circuit using PennyLane, Lines 1-2 are used to import the PennyLane and NumPy libraries. Line 3 is used to define the device parameters with wires and shots. Lines 5-8 is to define the circuit function which is used to generate the EPR pair circuit with the same Hadamard and CNOT gates. The output shows the number of times both qubits measured zero is same as one. 

\begin{tcolorbox}[colback=black!5!white,colframe=white!5!black,title=Output,width=8cm]
\begin{tiny}
\begin{verbatim}
>>result = circuit()
>>result.shape
(2, 1000)

>>np.all(result[0] == result[1])
True
    
\end{verbatim}
\end{tiny}
\end{tcolorbox}

% \subsection{Quantum Support Vector Machine}
%  Support Vector Machine (SVM) are a type of supervised learning algorithm which is used for classification. It is one of the most used classification machine learning algorithm. The Aim of SVM is to construct a maximum margin hyperplane which bifurcates the given data into two classes. The target of SVM is find out the maximum margin hyperplane i.e. a plane from which the support vectors are at maximal distance.

\subsection{An end-to-end tutorial: Quantum solution to max cut problem}
\subsubsection{Combinatorial Optimization problems}
Combinatorial Optimization Problems (COP) are a class of optimization problems where the domain of the objective function \textbf{F} discreet but large. Standard problems such as TSP, and max-cut problems are examples of such optimization problems. It can be considered as searching for an object out of a finite set of objects. A COP problem is specified with a bit string of length n and m clauses. Clauses are constraints on substring, which satisfies for a specific combination of binary string values. The objective function is defined for COP problems as Eq. \eqref{cop}: 
\begin{equation}
\label{cop}
    C(z) = \sum_{\alpha=1}^{m}C_\alpha(z)\\
\end{equation}
where $z = z_1z_2z_3....z_n$ \\
and $C_\alpha(z)=1$ if z satisfies the clause $\alpha$ and otherwise 0.

\subsubsection{Quantum Approximate Optimization Algorithm (QAOA)}
QAOA is specially designed and well-suited for solving combinatorial problems in NISQ devices. QAOA is a class of variational algorithms defined as parameterized unitary $U(\beta,\gamma)$. The parameters are used to define a quantum state $\ket{\psi(\beta,\gamma})$. Now, The QAOA algorithms target is to optimize these parameters as $\beta_{opt},\gamma_{opt}$ such that the designed quantum state reached its minimum energy which maps to the solution of the problem i.e $\ket{\psi(\beta_{opt},\gamma_{opt})}$ is the solution state. 

\subsubsection{Circuit representation as Hamiltonian Product}
The Hamiltonian: In QM, every physical system has a set of measurable properties, and each measurable property is associated with an operator. The operator associated with the system energy is known as the Hamiltonian. It is also used in VQE as minimizing the Hamiltonian can be mapped with some optimization/loss function minimization. The corresponding mathematical representation is provided in Eq. \eqref{eq:hami}.
\begin{equation}
\label{eq:hami}
        \lambda_{\text{min}} \le \lambda_{\theta} \equiv \langle \psi(\theta) |\Omega|\psi(\theta) \rangle
\end{equation}

\begin{figure}[hb]
    \centering
    \includegraphics[width=\linewidth]{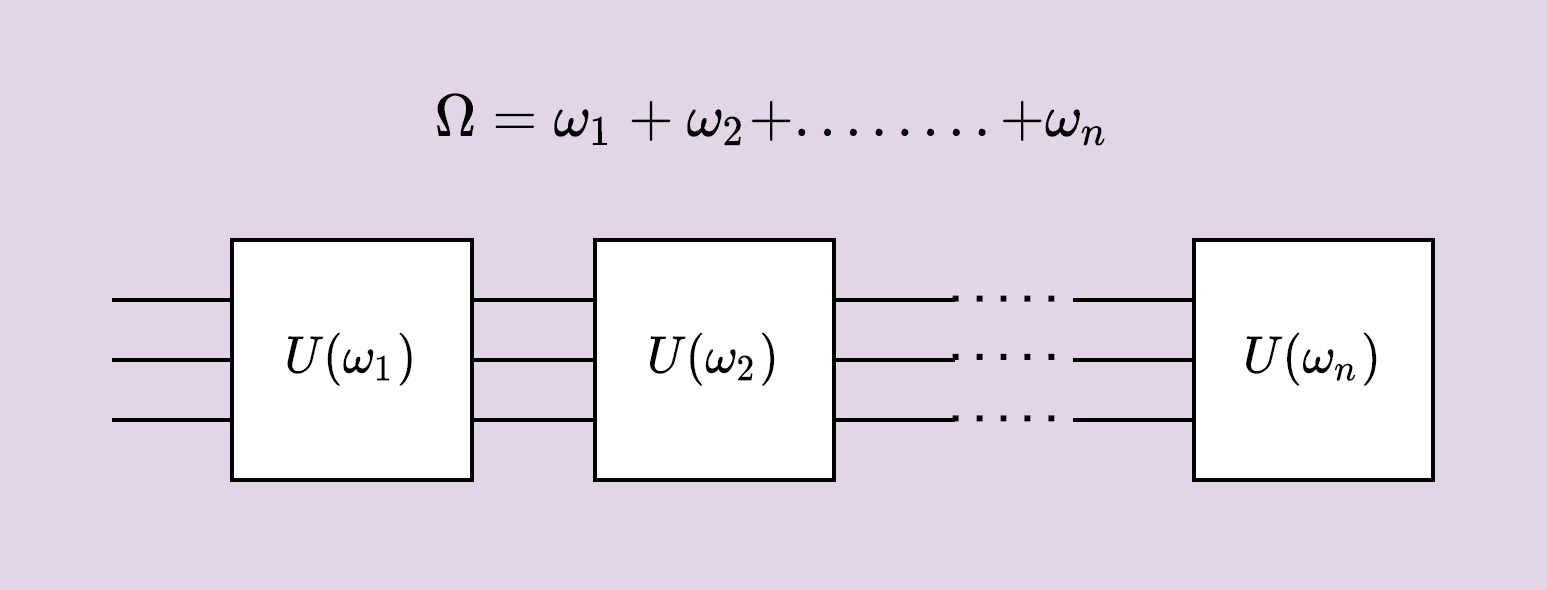}
    \caption{Hamiltonian Circuit }
    \label{fig:hamcir}
\end{figure}

Quantum Circuit: In the universal gate model QC, a circuit is represented as a sequence of quantum logic gates. The circuit can also be represented in terms of Hamiltonian, which is useful in several applications, including QAOA. The Gates can be represented as the time evolution Hamiltonian whose unitary transformation function is represented as Eq. \eqref{eq:ham} : 
\begin{equation}
\label{eq:ham}
    U(\Omega, \ t) \ = \ e^{\frac{-i\Omega t}{\hbar}}.
\end{equation}
where $\hbar$ is the reduced plank's constant. $\Omega$ is the Hamiltonian and $t$ represent the time.
It is hard to implement a quantum circuit for finding the exponential of Hamiltonian with many non-commuting terms, \textit{i.e.}, a Hamiltonian of the form:
$\Omega \ = \ \omega_1 \ + \omega_2 \ + \omega_3 \ + \ \cdots \ + \omega_n,$
is hard to implement, and an approximate unitary time evolution function is approximated using a method called Trotterization\cite{wiebe2010higher}  as given in Eq. \eqref{eq:toter}  :
\begin{equation}
\label{eq:toter}
    e^{A \ + \ B} \ \approx \ \Big(e^{A/n} e^{B/n}\Big)^{n}
\end{equation}
The approximate time evolution unitary is given as,  and the circuit is shown in Fig. \ref{fig:hamcir}:
\begin{equation}
U(\Omega, t, n) \ = \ \displaystyle\prod_{j \ = \ 1}^{n}
\displaystyle\prod_{k} e^{\frac{-i \omega_k t}{n}} \ \ \ \ \ \ \ \ \ \ \Omega \
= \ \displaystyle\sum_{k} \omega_k,    
\end{equation}
Coming back the QAOA discussion the Unitary $U(\beta,\gamma)$ is composed of two unitary as $U(\beta) = e^{-i\beta \Omega_{m}}$, and $U(\gamma) = e^{-i\gamma \Omega_{p}}$ where $\Omega_{m}$ is the mixer Hamiltonian and  $\Omega_{p}$ is the problem Hamiltonian. Now we will see how the state $\psi$ is prepared using the problem Hamiltonian and the mixer Hamiltonian (arranging in an alternating pattern).  as shown in Eq. \eqref{eq:8}
\begin{equation}
\label{eq:8}
    \ket{\psi(\beta,\gamma)} = \underbrace{U(\beta)U(\gamma)\cdots U(\beta)U(\gamma)}_{ptimes}\ket{\psi}
\end{equation}

% % Quantum Enhance Support Vector Machine (Q-SVM) 
% \url{https://pennylane.ai/qml/demos/tutorial_kernels_module.html}

% \url{https://learn.qiskit.org/course/machine-learning/quantum-feature-maps-kernels}

\subsubsection{Max Cut Problem}
\begin{figure}
    \centering
    \includegraphics[width=\linewidth]{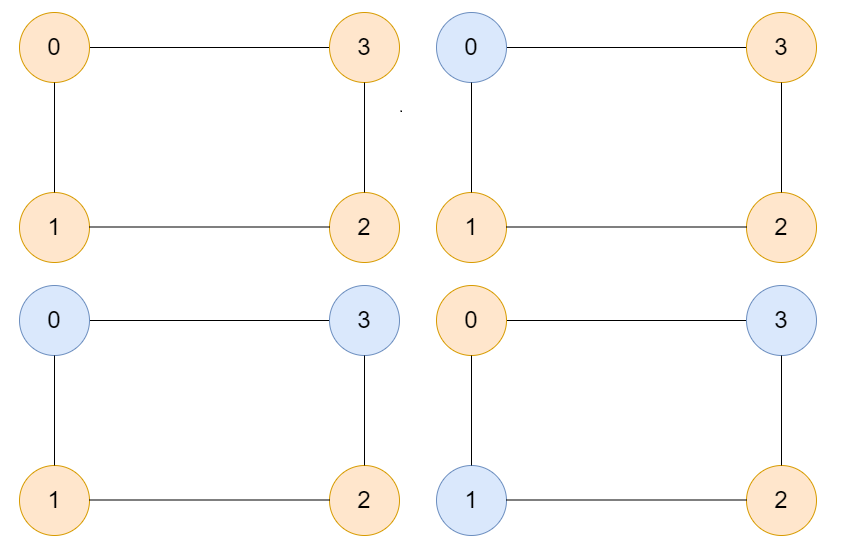}
    \caption{Max Cut Example: In the given example the corresponding number of edges for each cut is 0,2,2,4} 
    \label{fig:maxcut}
\end{figure}

Max cut is a COP problem in which a graph is partitioned into two sets of graph vertices under the criteria of maximizing the number of edges among two sets. As shown in Fig. \ref{fig:maxcut} with four max cut options out of $2^4=16$ combinations with 0,2,2,4 edges. As COP problems, the given example can be encoded with four length bit string where the bits can be mapped with $b_{3}b_{2}b_{1}b_{0}$ corresponds to $v_{3}v_{2}v_{1}v_{0}$ and solution state can be represented as 0101 or 1010 as per the Fig \ref{fig:maxcut}.
Formally a Maxcut problem can be defined as follows: Create a partition P($S_{1},S_{2}$) of the vertices over a Graph G (V, E); with V vertices, E edges and sets $S_{1},S_{2}$ such that it maximizes :
\begin{equation}
\Phi(P) = \sum_{\alpha=1}^{E}\Phi_\alpha(P)    
\end{equation}
where $\Phi$ represents the number of edge cut. $\Phi_{\alpha}(P)$ = 1 if P places one vertex from the $\alpha^{th}$ set edge in set $S_1$ and the other in set $S_2$, and $\Phi_{\alpha}(P)$ = 0 otherwise.

\subsection{The Solution: MAX CUT using QAOA }
In this section, it is described how Max-Cut is implemented with the help of quantum algorithms. but before looking into the implementation, let's see how a graph can be implemented in Python, which will be common for all three platforms. The networkx library provides the graph representation in Python language, which can be used to implement a graph as follows.

\begin{figure}
    \centering
    \includegraphics[width=\linewidth]{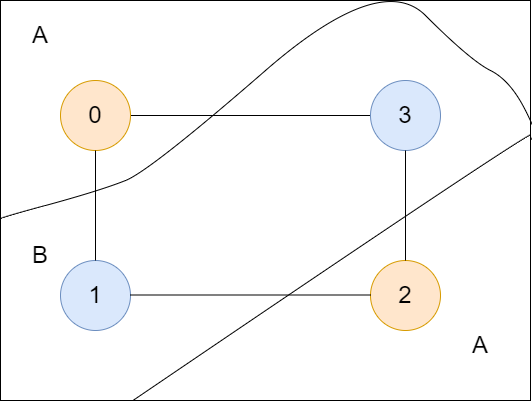}
    \caption{A Solution for the MAXCUT which shows two partitions A and B and 4 edge cuts}
    \label{fig:maxcutsol}
\end{figure}

\begin{tcolorbox}[width=8cm]
\begin{tiny}
\begin{verbatim}
import networkx as nx

G = nx.Graph()
G.add_nodes_from([0, 1, 2, 3])
G.add_edges_from([(0, 1), (1, 2), (2, 3), (3, 0)])
    
\end{verbatim}
\end{tiny}
\end{tcolorbox}

\subsubsection{Qiskit}
\begin{figure}
    \centering
    \includegraphics[width=\linewidth]{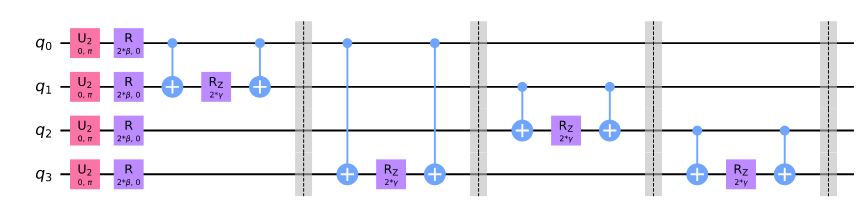}
    \caption{Max-Cut Circuit corresponding to the PennyLane implementation}
    \label{fig:maxcutcircuit}
\end{figure}

The QISKIT implementation of the MAX CUT Algorithm will follow the steps.
\begin{itemize}
    \item Step 1: Designing the mixer and problem Hamiltonian.   
    \item Step 2: Initial state may be the superposition of all qubits.
    \item Step 3: Creating the circuit as shown in Fig. \ref{fig:maxcutcircuit} by appending the mixer and problem Hamiltonian. 
    \item Step 4: Executing the circuit using a run time and optimizing the state function. The scikit optimizer is used for optimization.
\end{itemize}

% \onecolumn
% \begin{table}[]
%     \centering
%     \begin{tabular}{m{8cm}|m{8cm}}
%     \includegraphics[width=8cm]{testi.pdf}&
%     % &  \includegraphics[width=\linewidth]{testi.pdf}      
%     \end{tabular}
%     \caption{Caption}
%     \label{tab:my_label}
% \end{table}

% \twocolumn

\begin{tcolorbox}[colback=black!5!white,colframe=white!5!black,title= Max-Cut Implementation: Qiskit,width= 8.3cm]
% \textbf{Cirq Example of Bell State Creation}
% \begin{scriptsize}
\begin{tiny}
\begin{verbatim}
#library imports 
from qiskit import QuantumCircuit as QC
from qiskit import Aer, execute
from qiskit.circuit import Parameter
from scipy import minimize as min
#defining adjancency matrix for the given graph G
1.  adj_mat = nw.adjacency_matrix(G).todense()
# define qubits equal to the number of vertices
2.  qubits = 4 
# Code section to create a mixer unitary 
3.  b = Parameter("$\\beta$")
4.  unitary_mix = QC(nqubits)
5.  for i in range(0, qubits):
        unitary_mix.rx(2 * b, i)
## Code section to create a problem unitary
6. g = Parameter("$\\gamma$")
7. unitary_prob = QC(nqubits)
8. for p in list(G.edges()): 
9.     unitary_prob.rzz(2 * g, p[0], p[1])
10.    unitary_prob.barrier()

# Code section for the initial state 
##Initial state will be a superposition of all qubits
11. init = QC(qubits)
12. for i in range(0, qubits):
13.     init.h(i)

# Code section for circuit creation
# by appending mixer, problem, and initial circuit
14. circuit_maxcut = QC(qubits)
15. circuit_maxcut.append(init, [i for i in range(0, qubits)])
16. circuit_maxcut.append(unitary_prob, [i for i in range(0, qubits)])
17. circuit_maxcut.append(unitary_mix, [i for i in range(0, qubits)])

#code section for defining the objective function 
18. def obj_func(x, G):
19.     obj = 0
20.     for i, j in G.edges():
21.         if x[i] != x[j]:
22.             obj -= 1          
23. return obj

#code section for computing expectations
24. def exp_func(counts, G):
25.     av = 0
26.     s_count = 0
27.     for bitstring, count in counts.items():
28.     obj = obj_func(bitstring[::-1], G)
29.     av += obj * count
30.     s_count += count     
31. return avg/s_count
#code section for defining the optimizer circuit
32.  def create_qaoa_circ(G, theta):
33.     qubits = len(G.nodes())
34.     p = len(theta)//2  # number of alternating unitaries
35.     qc = QC(qubits)
36.     b = theta[:p]
37.     g = theta[p:]

# defining the initial_state applying Hadamard on all states
38.  for i in range(0, qubits):
39.     qc.h(i)
40.  for irep in range(0, p):

# Applying the problem unitary
41.  for p in list(G.edges()):
42.    qc.rzz(2 * g[irep], p[0], p[1])

# Applying the mixer unitary
43. for i in range(0, qubits):
44.      qc.rx(2 * b[irep], i)  
45. qc.measure_all()
46. return qc
# Function for executing the circuit
47. def get_expectation(G, shots=512):
48.     backend = Aer.get_backend('qasm_simulator')
49.     backend.shots = shots
50.     def execute_circ(theta):
51.     qc = create_qaoa_circ(G, theta)
52.     counts = backend.run(qc, seed_simulator=10, 
                 nshots=512).result().get_counts()
53.     return exp_func(counts, G)
54. return execute_circ
55. exp = get_expectation(G)
56. result = min(exp,[1.0, 1.0],method='COBYLA')
\end{verbatim}
\begin{footnotesize}
Source: \url{https://www.qiskit.org/}    
\end{footnotesize}
% \end{scriptsize}
\end{tiny} 
\end{tcolorbox}

Code Explanation:  The code consists of the tasks- importing required libraries, defining unitary (problem, mixer), defining an objective function, creating the circuit corresponding to the problem, and executing it for the optimized circuit to obtain the Max-Cut of the given graph. Lines (3-5) are used to define the mixer unitary and lines (6-10) are used to define the problem unitary. Lines (18-23) are used to define the objective function which is used to obtain the cut for the graph. Lines(47-54) are used to execute the circuit and the result is obtained in the results variable.

\subsubsection{PennyLane}
Implementing the MAX CUT problem using QAOA is much easier as compared to Qiskit due to its inherited support for QML problems. The implementation is provided in the textbox below.

\begin{tcolorbox}[colback=black!5!white,colframe=white!5!black,title= Max-Cut Implementation: PennyLane,width=8.3cm]
\begin{tiny}
% \begin{scriptsize}
\begin{verbatim}
#As earlier pl corresponds to pennylane library 
#pl_np corresponds to numpy from pennylane

# define wires equal to the vertices
1.  pl_np.random.seed()
2.  n_lines = 4

# Code section to create a mixer unitary 
3.  def unitary_mix(beta):
4.      for w in range(n_lines):
5.          pl.RX(2 * beta, wires=w)

# Code section to create a problem unitary 
6.    def unitary_prob(g):
7.      for e in G:
8.          line1 = e[0]
9.          line2 = e[1]
10.   pl.CNOT(wires=[line1, line2])
11.   pl.RZ(g, wires=line2)
12.   pl.CNOT(wires=[line1, line2])

#Code section for defining circuit
13. dev = pl.device("default.qubit", wires=n_lines, shots=1) 
14. @pl.qnode(dev)
15. def circuit(g, b, e=None, n_layers=1):
16.     for line in range(n_lines):
17.         pl.Hadamard(wires=line)
18.     for i in range(n_layers):
19.           unitary_prob(g[i])
20.           unitary_mix(b[i])
21.     if edge is None:
22. return pl.sample()
#code section for computing expectations in the optimization function 
23.  H = pl.PauliZ(e[0]) @ pl.PauliZ(e[1])
24.     return pl.expval(H)
25. def opt_circuit(n_layers=1):
# Create an initial state using a near-zero value
26. init_param =  
                  0.01 *pl_np.random.rand
                  (2, n_layers, requires_grad=True)
# Code section for defining the objective function
27. def obj_func(param):
28.     gammas = param[0]
29.     betas = param[1]
30.     neg_obj_func = 0
31.     for edge in G:
32.         neg_obj_func -= 0.5 * (1 - circuit(
                            gammas, betas, edge=edge,
                           n_layers=n_layers))
33. return neg_obj_func

# Intialize the optimizer function using ADAM optimizer
34. opt = pl.AdagradOptimizer(stepsize=0.5)
# Code section for parameter optimization
35.  param = init_param
36.  rep = 30
37.  for i in range(rep):
38.     param = opt.step(obj_func, param)
39.     if (i + 1) % 5 == 0:
40.       print("Obj_func after iteration {:5d}: 
            {: .7f}".format(i + 1, -objective(param)))
# sample measured bitstrings 500 times
41.   bit_strings = []
42.   n_samples = 500
43.   for i in range(0, n_samples):
44.         bit_strings.append
45.            (bitstring_to_int(circuit
46.                    (param[0], param[1], 
47.                    e=None, n_layers=n_layers)))
# Calculate optimal parameters and display it
48.      counts = pl_np.bincount(pl_np.array(bit_strings))
49.      most_freq_bit_string = pl_np.argmax(counts)
50.      print("Optimized (gamma, beta) vectors:
             \n{}".format(param[:, :n_layers]))
51.      print("Highest frequency bit string is: 
           {:04b}".format(most_freq_bit_string))
52. return -obj_func(param), bit_strings

# Apply the optimization of graph with p=1,2 and
# keep the bitstring1 and 2 will store the result
53. bitstring1 = opt_circuit(n_layers=1)[1]
54. bitstring2 = opt_circuit(n_layers=2)[1]
\end{verbatim}
\end{tiny}
\begin{footnotesize}
Source:\url{https://www.pennylane.ai/qml/demos/}    
\end{footnotesize}

% \end{scriptsize}
\end{tcolorbox}
Code Explanation:  The example code consists of the tasks- importing required libraries, defining the problem and mixer Hamiltonian, defining an objective function, creating the circuit corresponding to the problem, and executing it for the optimized circuit to obtain the Max-Cut of the given graph. Lines (3-5) are used to define the mixer Hamiltonian unitary and lines (6-12) are used to define the problem Hamiltonian unitary. Lines (27-33) are used to define the objective function which is used to obtain the cut for the graph. Lines (47-54) are used to execute the circuit and the result is obtained in the bitstring1, bitstring2 variable.

\section{Conclusion}
\label{sec:6}
In this exhaustive survey, the Quantum Software development tools including quantum simulators, and quantum cloud computers are discussed. The paper presented the survey cum tutorial that can be pretty helpful for beginners in quantum development. It is written in such a way that researchers from any background can benefit and easily understand all the requirements for quantum development. A detailed discussion, as well as comparative analysis, has been provided for Quantum development tools as well as hardware. The key point that distinguishes it from other surveys is that it not only surveys but also provides an end-to-end example code for different tools. The limitation of this survey is  that it fails to cover other interrelated areas, such as quantum communication, quantum cryptography, and quantum information processing, due to the incompatible scope of coverage. In the future, a similar kind of survey will be provided for other quantum fields as well. 

\bibliographystyle{IEEEtran}
\bibliography{export.bib}

\end{document}